\documentclass[a4paper,11pt]{article}
\pdfoutput=1
\usepackage{jheppub}
\usepackage{amsmath,amssymb,bm,graphicx,bbold,epsf,colordvi}
\usepackage{lipsum}
\usepackage{soul}
\usepackage{braket}
\allowdisplaybreaks 
\usepackage{color}
\newcommand\f{\frac}
\newcommand\as{\alpha_s}
\newcommand{\ba}{\begin{eqnarray}}
\newcommand{\ea}{\end{eqnarray}}
\def\bea#1\eea{\begin{align}#1\end{align}}
\newcommand{\bef}{\begin{figure*}[t]\centering}
\newcommand{\eef}{\end{figure*}}
\newcommand{\be}{\begin{equation}}
\newcommand{\ee}{\end{equation}}
\newcommand{\nn}{\nonumber}

\def\OMIT#1{{}}

\def\nslash{n\!\!\!\slash}

\begin{document}

\title{The soft drop groomed jet radius at NLL}

\author[a,b]{Zhong-Bo Kang,}
\author[c,d]{Kyle Lee,}
\author[e]{Xiaohui Liu,}
\author[f]{Duff Neill}
\author[g,h]{and Felix Ringer}

\affiliation[a]{Department of Physics and Astronomy, University of California, Los Angeles, CA 90095, USA}
\affiliation[b]{Mani L. Bhaumik Institute for Theoretical Physics, University of California, Los Angeles, CA 90095, USA}
\affiliation[c]{C.N. Yang Institute for Theoretical Physics, Stony Brook University, Stony Brook, NY 11794, USA}
\affiliation[d]{Department of Physics and Astronomy, Stony Brook University, Stony Brook, NY 11794, USA}
\affiliation[e]{Center of Advanced Quantum Studies, Department of Physics, Beijing Normal University, Beijing 100875, China}
\affiliation[f]{Theoretical Division, MS B283, Los Alamos National Laboratory, Los Alamos, NM 87545, USA}
\affiliation[g]{Physics Department, University of California, Berkeley, CA 94720, USA}
\affiliation[h]{Nuclear Science Division, Lawrence Berkeley National Laboratory, Berkeley, CA 94720, USA}
                                   
\emailAdd{zkang@physics.ucla.edu}
\emailAdd{kunsu.lee@stonybrook.edu}
\emailAdd{xiliu@bnu.edu.cn}
\emailAdd{duff.neill@gmail.com}
\emailAdd{fmringer@berkeley.edu}


\abstract{We present results for the soft drop groomed jet radius $R_g$ at next-to-leading logarithmic accuracy. The radius of a groomed jet which corresponds to the angle between the two branches passing the soft drop criterion is one of the characteristic observables relevant for the precise understanding of groomed jet substructure. We establish a factorization formalism that allows for the resummation of all relevant large logarithms, which is based on demonstrating the all order  equivalence to a jet veto in the region between the boundaries of the groomed and ungroomed jet. Non-global logarithms including clustering effects due to the Cambridge/Aachen algorithm are resummed to all orders using a suitable Monte Carlo algorithm. We perform numerical calculations and find a very good agreement with Pythia 8 simulations. We provide theoretical predictions for the LHC and RHIC.}

\maketitle

\section{Introduction~\label{sec:intro}}

At present day collider experiments such as the LHC and RHIC, highly energetic jets play an important role as precision probes of the Standard Model and beyond. In the past years, jet substructure techniques have become important tools in high energy particle and nuclear physics. One of the important techniques that have been developed is jet grooming which is designed to remove soft wide-angle radiation from the identified jets. Algorithms that remove the soft contamination of jets allow for a more direct comparison of perturbative QCD calculations and data due to the reduced sensitivity to nonperturbative effects. Different grooming algorithms have been developed in the literature such as~\cite{Krohn:2009th,Ellis:2009me,Dasgupta:2013ihk,Larkoski:2014wba}. In this work, we focus on the soft drop grooming algorithm of~\cite{Larkoski:2014wba}. Both on the experimental~\cite{Aaboud:2017qwh,Sirunyan:2018xdh,Aaboud:2019aii,ATLAS:2019sol} and the theoretical side~\cite{Larkoski:2017jix,Asquith:2018igt,Frye:2016aiz,Marzani:2017mva,Kang:2018jwa,Kang:2018vgn,Larkoski:2017cqq,Makris:2017arq,Hoang:2017kmk,Baron:2018nfz,Makris:2018npl,Kardos:2018kth,Napoletano:2018ohv,Chien:2019osu,Lee:2019lge,Gutierrez-Reyes:2019msa}, significant progress has been made recently in improving our understanding of soft drop groomed jet observables. In the heavy-ion community, soft drop groomed jet substructure observables have also received increasing attention from both experiment~\cite{Sirunyan:2017bsd,Sirunyan:2018gct, Acharya:2019djg,Kauder:2017cvz} and theory~\cite{Mehtar-Tani:2016aco,Chien:2016led,Milhano:2017nzm,Chang:2017gkt,Li:2017wwc,Chien:2018dfn,Sirimanna:2019bgl,Caucal:2019uvr,KunnawalkamElayavalli:2017hxo,Casalderrey-Solana:2019ubu,Ringer:2019rfk}. Jet grooming techniques can be used to isolate different aspects of jet quenching and may help to discriminate between different model assumptions~\cite{Andrews:2018jcm}.

One of the interesting features of soft drop grooming is that the radius of the groomed jet is adjusted dynamically, capturing only the hard collinear core of the jet~\cite{Larkoski:2014wba} which we study in this work within perturbative QCD. We consider inclusive jet production $pp\to{\rm jet}+X$ where jets are identified with the anti-$k_T$ algorithm~\cite{Cacciari:2008gp} with a given radius $R$. Following the soft drop algorithm, the identified jets are then reclustered with the Cambridge/Aachen (C/A) algorithm~\cite{Dokshitzer:1997in,Wobisch:1998wt}. The obtained angular ordered clustering tree is then declustered recursively where at each step the soft drop condition is checked 
\bea
\label{eq:sd}
\f{{\rm min}[p_{T1},p_{T2}]}{p_{T1}+p_{T2}} > z_{\rm cut}\left(\f{\Delta R_{12}}{R}\right)^\beta \,.
\eea
Here $p_{T1,2}$ are the transverse momenta of the two branches obtained at each declustering step and $\Delta R_{12}^2=\Delta\eta^2+\Delta\phi^2$ is their geometric distance. Soft branches that fail the criterion are removed from the jet. The algorithm terminates when the criterion is met and the particles in the remaining two branches constitute the groomed jet. The soft threshold $z_{\rm cut}$ and the angular exponent $\beta$ are fixed parameters that determine how aggressively soft radiation is removed. For $\beta=0$, the soft drop algorithm reduces to the modified mass drop tagger (mMDT)~\cite{Dasgupta:2013ihk}. Two variables that characterize important features of the soft drop groomed jet are the momentum sharing fraction $z_g$ and the groomed jet radius $R_g$. Their values are obtained from the kinematics of the two remaining branches when the soft drop algorithm terminates
\bea
z_g= \f{{\rm min}[p_{T1},p_{T2}]}{p_{T1}+p_{T2}}\,,\qquad 
R_g=\Delta R_{12} = R \,\theta_g \,.
\eea
Often the variable $\theta_g$ is used which corresponds to the geometric distance of the two remaining branches normalized by the radius $R$ of the ungroomed initial jet. Note that unlike the external parameter $R$ which is the radius of the initial jet, the groomed radius $R_g$ is a distribution which is determined through the soft drop grooming procedure. Since the C/A algorithm first clusters particles that are closer in angle, the groomed jet radius $R_g$ defines the maximally allowed angle between two branches that can be clustered. Similar to the radius $R$ of the initial jet, the distance $R_g$ constitutes the radius of the soft drop groomed jet. By analyzing the active area of recursive $k_T$-type algorithms, it was found in~\cite{Cacciari:2008gn} that jets have an area of the order ${\cal O}(\pi R^2)$. A similar analysis was performed for the groomed radius $R_g$ in~\cite{Larkoski:2014wba} verifying that the active area of a soft drop groomed jet is of the order~${\cal O}(\pi R_g^2)$.

In the phenomenologically relevant limit of $R_g\ll 1$ and $z_{\rm cut}\ll 1$, large logarithms may spoil the convergence of the perturbative series expansion in terms of the QCD strong coupling constant. In~\cite{Larkoski:2014wba}, the soft drop groomed radius was calculated within the modified leading-logarithmic (MLL) approximation. In this work, we extend the calculation to next-to-leading logarithmic (NLL) accuracy using a factorization formalism developed within Soft Collinear Effective Theory (SCET)~\cite{Bauer:2000ew, Bauer:2000yr, Bauer:2001ct, Bauer:2001yt,Beneke:2002ph}, which is suitable for the extension to yet higher perturbative accuracy. Besides the resummation of logarithms of $R_g$, we also take into account logarithms of $R$ and the soft threshold parameter $z_{\rm cut}$. Based on the equivalence of the $R_g$ measurement and a jet veto on emissions between the splitting that satisfies the soft drop criterion and the boundary of the initial ungroomed jet, we establish the all order factorization framework. Different than for example the groomed jet mass distribution, non-global logarithms (NGLs)~\cite{Dasgupta:2001sh} directly contribute to the cross section starting at NLL accuracy. The use of the C/A algorithm introduces clustering constraints that give rise to clustering logarithms associated with both NGLs as well as global logarithms, which are referred to as Abelian clustering logarithms~\cite{Delenda:2006nf,KhelifaKerfa:2011zu,Delenda:2012mm,Dasgupta:2012hg,Kelley:2012kj,Kelley:2012zs}. We resum the NGLs including clustering constraints and the Abelian clustering logarithms at leading logarithmic (LL) accuracy and leading color using a suitable Monte Carlo algorithm which we introduce here following the work of~\cite{Dasgupta:2001sh,Appleby:2002ke,Neill:2018yet}. 

The remainder of this paper is organized as follows. In section~\ref{sec:2}, we outline the factorization formalism developed in this work based on the equivalence between the groomed radius measurement and a jet veto when $R_g \ll 1$ and $z_{\rm cut} \ll 1$. We identify the relevant NGLs and Abelian clustering logarithms and perform the relevant fixed order calculations. In section~\ref{sec:3}, we introduce the Monte Carlo setup that allows for the all order resummation of NGLs and clustering logarithms at LL accuracy needed to achieve the overall accuracy at NLL. Numerical studies and a comparison to Pythia 8 simulations are presented in section~\ref{sec:4}. We draw our conclusions in section~\ref{sec:5} and present an outlook.

\section{Factorization and resummation \label{sec:2}}

In this section, we develop the factorization theorem for the soft drop groomed jet radius within SCET. We start from the cross section $\Sigma(\theta_g)$ differential in the transverse momentum $p_T$ and rapidity $\eta$ of the observed jet, but cumulative in the groomed jet radius where any value below $\theta_g$ contributes. The distribution differential in $\theta_g$ can then be obtained as
\bea\label{eq:diff}
\frac{\mathrm{d}\sigma}{\mathrm{d}\eta \,\mathrm{d}p_T \,\mathrm{d}\theta_g} =  \frac{\mathrm{d}}{\mathrm{d}\theta_g} \frac{\mathrm{d}\Sigma(\theta_g)}{\mathrm{d}\eta\, \mathrm{d}p_T} \,.
\eea
We work in the limit where the observed jet is sufficiently collimated $R\ll 1$ and we drop power corrections of the form ${\cal O}(R^2)$. This type of power corrections are generally found to be small even for relatively large values of the jet radius~\cite{Mukherjee:2012uz}. In this limit, the production of an energetic parton in a hard-scattering event factorizes from the formation and evolution of the jet initiated by the produced parton. The hard-scattering process $ab\to c$ is described by hard functions $H_{ab}^c$ which are known analytically to next-to-leading order (NLO)~\cite{Aversa:1988vb,Jager:2002xm}. The subsequent formation and evolution of the jet is described by a semi-inclusive jet function ${\cal G}_c$~\cite{Catani:2013oma,Dasgupta:2014yra,Kaufmann:2015hma,Kang:2016mcy,Dai:2016hzf}. This separation is generally expected to hold to all orders due to the universality of the collinear limit in QCD~\cite{Elder:2017bkd}. We can thus write the cumulative cross section in $\theta_g$ as
\bea\label{eq:inclusive}
 \f{\mathrm{d}\Sigma(\theta_g)}{\mathrm{d}p_T\, \mathrm{d}\eta} = \sum_{abc} f_a(x_a,\mu) \otimes f_b(x_b,\mu) \otimes H_{ab}^c(x_a,x_b,\eta,p_T/z,\mu)\otimes {\cal G}_c(z,\theta_g,p_T R, \mu; z_{\rm cut}, \beta )\;,
\eea
where $f_{a,b}$ denote the parton distribution functions (PDFs) for finding partons $a,b$ in the colliding protons. Here, $\otimes$ denote appropriate integrals over the longitudinal momentum fractions $x_{a,b}$ of the initial partons and $z$ which is the fraction of transverse momentum contained in the observed jet relative to the scattered parton $c$. Note that the jet rapidity $\eta$ only appears in the hard functions $H_{ab}^c$ when subleading terms $\sim {\cal O}(R^2)$ are ignored. On the other hand, the entire dependence on $\theta_g$ and the grooming parameters is contained in the jet function ${\cal G}_c$. Single logarithms of the jet radius $\alpha_s^n\ln^n R$ can be resummed by solving the renormalization group (RG) evolution equation (DGLAP) associated with the jet function ${\cal G}_c$ which is given by
\be
\mu\f{\mathrm{d}}{\mathrm{d}\mu}{\cal G}_c = \f{\as}{2\pi} \sum_{d} P_{dc}\otimes {\cal G}_d \,.
\ee
Here, $P_{dc}$ denote the Altarelli-Parisi splitting functions which can be computed order by order in $\alpha_s$. In the kinematic region where $z_{\rm cut} \sim \theta_g\sim {\cal O}(1)$, the factorization theorem in eq.~(\ref{eq:inclusive}) is sufficient to carry out calculations at fixed order in perturbation theory. In the phenomenologically relevant region where $z_{\rm cut}\ll 1$ and $\theta_g\ll 1$, logarithms of the form $\alpha_s^n \ln^{2n} \theta_g$ ($\beta>0$) and $\alpha_s^n\ln^{2n} z_{\rm cut}$ may spoil the perturbative convergence and an all order resummation is required. This can be achieved by a refactorization of the semi-inclusive jet function ${\cal G}_c$ in order to separate the physics at different scales in the relevant kinematic regime. The associated RG evolution equations then allow for the resummation of all relevant large logarithms.

\subsection{Refactorization of the semi-inclusive jet function}

\begin{figure}[t]\centering
\includegraphics[width=3.3in]{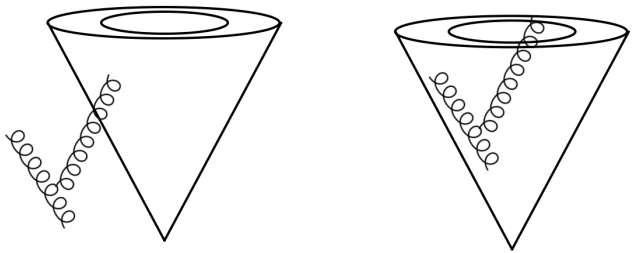} 
\caption{Configurations that give rise to different NGL contributions at NNLO $\sim\alpha_s^2\ln^2 z_{\rm cut}$ (left) and $\sim\alpha_s^2\ln^2(z_{\rm cut}\theta_g^\beta)$ (right). The inner ellipse denotes the area set by the groomed radius $R_g$ inside the original ungroomed jet.~\label{fig:correlation}}
\end{figure}

In this section we discuss the refactorization of the semi-inclusive jet function ${\cal G}_c$ in the limit when both $z_{\rm cut}\ll 1$ and $\theta_g\ll 1$. We make use of power counting arguments to establish the refactorization. First, we consider energetic collinear radiation at the jet scale $\mu_{\cal H}\sim p_T R$. To NLO, these are given by out-of-jet radiation diagrams, see for example~\cite{Kang:2017mda,Kang:2017glf}. The scaling of the associated collinear mode in terms of light-cone momentum components is given by
\bea\label{eq:hard}
&  p_{\cal H} = ( p^-,p^+,p^\perp ) \sim  p_T(1,R^2, R)\,.
\eea
Second, we consider soft modes that describe wide angle soft radiation\footnote{Although this mode is both collinear and soft, we just refer to it as soft since it would correspond to soft radiation when boosted to a frame where the in-jet and out-of-jet region are complementary hemispheres.} within the jet at an angle $\theta\sim R$. If such radiation passes grooming with momentum fraction $z > z_{\rm cut}$, then the scaling $\theta_g \ll 1$ would be violated. Therefore, this kind of soft radiation must fail the grooming condition and it is thus independent of the $\theta_g$ measurement. The associated momentum scaling is
\bea\label{eq:soft_graway}
&  p_s^{\notin \text{gr}}  \sim z_{\rm cut} p_T(1,R^2, R) \,.
\eea
The superscript indicates that the soft radiation considered here fails the grooming condition. The radiation associated with the two modes identified so far are taken into account by two functions, ${\cal H}^n_{c\to i}$ and $S_{i,n}^{\notin \text{gr}}$. Both are independent of the measured groomed jet radius and the same modes were obtained in other factorization theorems of groomed jet substructure observables before, see for example~\cite{Frye:2016aiz,Kang:2018jwa,Kang:2018vgn}. At this point, we obtain the following refactorized expression of the semi-inclusive jet function
\begin{align}
\label{eq:refactorize}
{\cal G}_c(z,\theta_g,p_T R, \mu; z_{\rm cut}, \beta ) =& \sum_{i=q,\bar q,g} \sum_n \mathcal{H}^n_{c\to i}(z,p_T R,\mu) \nn\\
&\otimes_{\Omega}\, S^{\notin \text{gr}}_{i,n}\left(z_{\rm cut}p_T R,\mu;\beta \right) \, 
\mathcal{F}_i(\theta_g,p_T R,\mu;z_{\rm cut},\beta)  \,.
\end{align}
Here the additional summation over $n$ and $\otimes_\Omega$ are introduced to account for NGLs~\cite{Becher:2015hka,Larkoski:2015zka} as discussed in more detail below. The remaining function $\mathcal{F}_i$ contains the dependence on the groomed radius $\theta_g$. Here we need to consider both collinear and collinear-soft radiation~\cite{Bauer:2011uc}. The collinear radiation with momentum fraction $z\sim 1$ always passes the grooming condition at leading power. The collinear-soft radiation instead is sensitive to the grooming condition and has $z\sim z_{\rm cut} \theta_g^\beta\ll 1$. In both cases, the characteristic angular scale is $\theta\sim R_g$ and the radiation described by $\mathcal{F}_i$ is thus insensitive to the boundary of the initial ungroomed jet. Note that this situation is different than for example the mode decomposition when the jet mass is measured to be small $m_J^2/p_T^2\ll 1$. In that case, the angle of the collinear and collinear-soft radiation is set by $\theta\sim\sqrt{m_J^2/p_T^2/z}$ which depends on the scaling of the different momentum fractions. Because of this scaling that is imposed by the small jet mass measurement, the collinear and the collinear-soft radiation can be treated as two independent sectors. For the soft drop groomed jet radius, we thus have two additional modes with the following momentum scalings
\bea\label{eq:scalings}
&p_c  \sim p_T(1,R_g^2, R_g) \,, \\
&p_s^\text{gr} \sim  z_{\rm cut} p_T \left(\frac{R_g}{R}\right)^\beta(1,R_g^2, R_g)\,.
\eea
The soft drop declustering algorithm makes a further separation of these two modes to all orders highly non-trivial. However, as will be demonstrated in the next section~\ref{sec:proof}, there is a formal equivalence between the soft drop declustering algorithm and a jet veto procedure when $\theta_g$ is measured to be small. We can treat the groomed jet with radius $R_g$ as the signal jet and the collinear-soft branches are subject to a veto condition where the veto parameter is set to $z_{\rm cut}\theta_g^\beta p_T$. With this equivalence we can further refactorize $\mathcal{F}_i$ in eq.~(\ref{eq:refactorize}) using results from jet veto calculations, see for example~\cite{Becher:2015hka, Banfi:2002hw, Banfi:2010pa,Liu:2012sz,Liu:2013hba}.
We find that we can write $\mathcal{F}_i$ in terms of a collinear function $C_i$ and a collinear-soft function $S_i^{\in \text{gr}}$ as
\bea\label{eq:refactorize2}
\mathcal{F}_i(\theta_g,p_T R,\mu;z_{\rm cut},\beta)\, &= \,\sum_m  C_{i}^m\left( \theta_g\, p_T R,\mu\right) 
\otimes_{\Omega} S^{\in \text{gr}}_{i,m}(z_{\rm cut} \theta_g^{1+\beta}\, p_T R,\mu;  \beta )  \,. 
\eea
NGLs are accounted for by the convolution integrals denoted by $\otimes_{\Omega}$ and the additional sum over the directions of collinear emissions $m$. Here we follow the notation introduced in~\cite{Becher:2015hka}, see also eq.~(\ref{eq:refactorize}) above. Collinear final-state particles set the directions for a multi-Wilson line structure. We sum over these directions $n,m$ in eqs.~(\ref{eq:refactorize}) and~(\ref{eq:refactorize2}) and $\otimes_\Omega$ indicates that angular integrals cannot be carried out independently which gives rise to correlations between the different functions resulting in NGLs. 
The NGLs in $z_{\rm cut}$ associated with the functions $\mathcal{H}_{c\to i}$ and $S_i^{\notin \text{gr}}$ in eq.~(\ref{eq:refactorize}) will affect the $\theta_g$ distribution only indirectly through the relative normalization of partonic channels. We note that the contribution from the correlation between the $\theta_g$ sensitive and insensitive modes are power suppressed~\cite{Larkoski:2014wba}. This can also be seen from eq.~(\ref{eq:refactorize}), where $\mathcal{H}_{c\to i}$ and $S^{\in \text{gr}}_i $ are fully decoupled from $\mathcal{F}_i$. In addition, beyond NLO clustering logarithms need to be taken into account due to the mismatch between the grooming operation acting on branches rather than individual partons and the use of the C/A algorithm. These contributions appear either in the soft function $S_i^{\notin \text{gr}}$ and the combination of $C_i \otimes_{\Omega} S^{\in \text{gr}}_i $. Due to the summation over the collinear emission history and the angular convolution structure, the analytical resummation using the refactorized cross section is usually difficult and the approaches discussed in the literature typically resort to the Monte Carlo methods~\cite{Dasgupta:2001sh,Appleby:2002ke,Banfi:2002hw,Becher:2015hka,Neill:2018yet}. Up to NLL using the known jet veto results~\cite{Banfi:2010pa, Liu:2012sz, Tackmann:2012bt, Dasgupta:2001sh, Delenda:2006nf, Delenda:2012mm}, we can write $\mathcal{F}_i$ as 
\begin{align}
\label{eq:NLL}
\mathcal{F}_i(\theta_g,p_T R,\mu;z_{\rm cut},\beta) =&\, \langle  C_i( \theta_g\, p_T R,\mu) \rangle \, \langle S^{\in \text{gr}}_i (z_{\rm cut} \theta_g^{1+\beta}\, p_T R,\mu;  \beta)  \rangle
\nn\\ 
&\times\, {\cal S}_{i,{\rm NGL}}^{\rm C/A}(t,\theta_g) \,
  \mathcal{A}_{i,{\rm Abel.}}^{\rm C/A}(t,\theta_g) \,. 
\end{align}
Here $\langle \dots \rangle$ indicates that we performed the solid angle integration, which thus allows us to solve the RG evolution equations of the collinear and collinear-soft function analytically. Here we define the variable $t$ as
\bea\label{eq:t}
t = \frac{1}{2\pi} \int_{ z_{\rm cut} \theta_g^\beta p_T}^{p_T} \frac{\mathrm{d} k_T}{k_T }
\alpha_s(k_T) \,. 
\eea
The NGLs due to the correlation of the radiation near the boundary of the groomed jet in eq.~(\ref{eq:NLL}) are taken into account by the function ${\cal S}^{\rm C/A}_{i,{\rm NGL}}(t,\theta_g)$ which has the following perturbative expansion
\bea\label{eq:NGLstructure}
{\cal S}^{\rm C/A}_{i,{\rm NGL}}(t,\theta_g) = 1 + \sum_{n=2} S^{\rm C/A}_{i,n}(\theta_g)\, t^n \,,
\eea
with coefficients $S^{\rm C/A}_{i,n}(\theta_g)$. The relevant configuration at NNLO is illustrated on the right side of Fig.~\ref{fig:correlation}, which takes into account correlations between emissions inside and outside the groomed jet but inside the initial ungroomed jet boundary. At NNLO, these NGLs are of the form $\sim\alpha_s^2\ln^2(z_{\rm cut}\theta_g^\beta)$. As the phase space of the in-and-out configurations is affected by the C/A algorithm, the numerical size of the NGLs are reduced due to clustering effects. The C/A algorithm also introduces global Abelian logarithms. The function ${\cal A}^{\rm C/A}_{i,{\rm Abel.}}(t,\theta_g)$ takes into account this contribution which can be calculated perturbatively as
\bea\label{eq:ABstructure}
{\cal A}^{\rm C/A}_{i,{\rm Abel.}}(t,\theta_g) = 1 + \sum_{n=2} A^{\rm C/A}_{i,n}(\theta_g)\, t^n \,,
\eea
with coefficients $A^{\rm C/A}_{i,n}(\theta_g)$. In sections~\ref{sec:ngl} and~\ref{sec:ab}, we compute the leading NGLs and Abelian clustering logarithms at NNLO and determine the coefficients $S^{\rm C/A}_{i,2}$ and $A^{\rm C/A}_{i,2}$. In order to achieve the resummation at LL and leading color, the Abelian clustering logarithms and NGLs are captured simultaneously by a suitable Monte Carlo algorithm as discussed in section~\ref{sec:3}. To leading logarithmic accuracy, the NGLs resulting due to the correlation of radiation inside and outside of the initial ungroomed jet can be resummed using the Monte Carlo results of~\cite{Dasgupta:2001sh}. Clustering corrections to this class of NGLs are power suppressed as the ungroomed jet is identified with the anti-$k_T$ algorithm. The respective NNLO configuration $\sim\alpha_s^2\ln^2 z_{\rm cut}$ is illustrated on the left side of Fig.~\ref{fig:correlation}. To NLL accuracy, we can thus write the convolution structure in eq.~(\ref{eq:refactorize}) as
\bea
\mathcal{H}^n_{c\to i}(z,p_T R,\mu) &\otimes_{\Omega}\, S^{\notin \text{gr}}_{i,n}\left(z_{\rm cut}p_T R,\mu;\beta \right)\;\to& \nn \\ &\langle \mathcal{H}_{c\to i}(z,p_T R,\mu) \rangle\, \langle S^{\notin \text{gr}}_{i}\left(z_{\rm cut}p_T R,\mu;\beta \right)\rangle\, {\cal S}_{i,{\rm NGL}}(t',z_{\rm cut}) \,.
\eea
Here $t'$ is defined as $t$ in eq.~(\ref{eq:t}), but with the lower integration limit replaced by $z_{\rm cut}p_T$. Therefore, $\, {\cal S}_{\rm NGL}(t',z_{\rm cut})$ can be obtained directly from the Monte Carlo result in~\cite{Dasgupta:2001sh} to leading logarithmic accuracy and leading color.

In order to realize the resummation at NLL accuracy, all components of the refactorized semi-inclusive jet function need to be calculated to NLO. The hard-collinear matching coefficients $\mathcal{H}_{c\to i}$ at NLO can be found in~\cite{Kang:2017mda,Kang:2017glf}. The operator definition of the soft function $S^{\notin \text{gr}}_i$ and its result at NLO can be found in~\cite{Kang:2018jwa}. Both functions do not directly affect the shape of the $\theta_g$ distribution, but they are important in order to determine the fractions of quark and gluon jets. The operator definitions of the remaining functions $C_i$ and $S^{\in \text{gr}}_i$ that appear in eq.~\eqref{eq:refactorize2} can be readily obtained by including the soft drop grooming operation in the relevant functions, see~\cite{Becher:2015hka}. In this work, we calculate the functions $\langle C_i \rangle$ and $\langle S^{\in \text{gr}}_i \rangle$ in the refactorized expression in eq.~(\ref{eq:NLL}) to NLO in order to achieve the resummation at NLL. The operator definitions and the results at NLO are presented in sections~\ref{sec:collinearfunction} and~\ref{sec:collinearsoftfunction} below.

\subsection{Equivalence between the soft drop groomed radius and the jet veto case~\label{sec:proof}}

In this section, we show the equivalence between the soft drop declustering algorithm and a jet veto on emissions outside the groomed jet for small values of the groomed radius $\theta_g \ll 1$ and $z_{\rm cut} \ll 1$, in which case the collinear and collinear-soft modes are well-defined. Non-trivial examples of this equivalence at NNLO are presented in Appendix~\ref{sec:NNLO}. 

The measurement functions of the collinear and the soft sectors are identical for both cases and we thus focus only on the collinear-soft radiation. We denote branches with collinear-soft scaling in the C/A clustering tree by $J_i$ which need to be tested against the soft drop criterion. The energetic collinear branch is denoted by $J$ which can also contain further collinear-soft radiation and $J$ by itself is not necessarily the final groomed jet. It is sufficient to consider a single collinear branch as two collinear branches always pass the soft drop condition. Due to angular ordering the collinear-soft branches $J_i$ are not clustered together, i.e. $\theta_{J_i,J_j} > \theta_{J_{i(j)},J}$ for all $i$ and $j$. Here we use the notation $\theta_{a,b}^2 = \eta_{a,b}^2 + \phi_{a,b}^2$ as the angular distance between the branches $a$ and $b$. Let us first consider the case of one collinear-soft branch $J_1$ for the cumulative distribution of $R_g$. The corresponding measurement function can be written as
\bea
{\cal M}_1(J_1) & = \Theta(\theta_{J_1,J} < R_g)\Theta(J_1 p) + \Theta(J_1 f) \nn \\
& \equiv {\cal M}_1(J_1 p) +  {\cal M}_1(J_1 f) \,,
\eea
where ``$p$" (``$f$'') means that $J_1$ passes (fails) the soft drop criterion. More specifically, 
\bea
\Theta(J_1 p ) = \Theta\left(p_{T,J_1}  - z_{\rm cut}  
\bigg(\frac{\theta_{J_1J}}{R}\right)^\beta p_{T,J}\bigg)\,,
\eea 
and 
\bea 
\Theta(J_1 f ) = \Theta\left(z_{\rm cut}  
\bigg(\frac{\theta_{J_1J}}{R}\right)^\beta p_{T,J}
- p_{T,J_1}\bigg) \,,
\eea
where we have used the fact that $p_{T,J}+p_{T,J_1} \approx p_{T,J}$ at leading power. One can directly see the equivalence between the soft drop procedure and the jet veto when there is only one collinear-soft branch since the measurement function can be written as
\bea
{\cal M}_1(J_1) =  \Theta(\theta_{J_1,J}<R_g)  +  \Theta(\theta_{J_1,J}>R_g) \Theta(J_1 f) \,. \qquad
\eea
If the separation of the collinear-soft branch $J_1$ from the collinear one $J$ is larger than $R_g$, i.e. outside the ``signal jet", the branch is required to be below the jet veto threshold, in this case $ z_{\rm cut}  \theta_{J_1,J}^\beta p_{T,J}$. On the other hand, if $\theta_{J_1,J}$ is less than $R_g$, $J_1$ is within the energetic signal jet and thus will always be kept. To proceed, we first note that
\bea
{\overline {\cal M}}_1 \equiv
1 -  {\cal M}_1=   \Theta(\theta_{J_1,J} > R_g)\Theta(J_1 p) \,,
\eea
which requires $\theta_{J_1,J} > R_g$. More generally for multiple branches, the measure 
$1 -  \prod_i^N{\cal M}_1(J_i) $ will require $\theta_{J_i,J} > R_g$ for at least one of the $J_i$ with $i=1 \dots N$. For 2 collinear-soft branches, the measurement function is given by
\bea\label{eq:M2}
{\cal M}_2  = & \sum_{\text{perm}}\, \Theta(J_2)\left[ {\cal M}_1(J_1) \,   {\cal M}_1(J_2 f)
+ \, {\cal M}_1(J_2 p) \right] \nn \\
=& \sum_{\text{perm}} \,  \Theta(J_2)\left[ {\cal M}_1(J_1)  \,  {\cal M}_1(J_2)
+ \, {\cal M}_1( J_2 p ) {\overline {\cal M}}_1  \right] \nn \\
=& \, {\cal M}_1(J_1)  \,  {\cal M}_1(J_2) \,,
\eea
where we introduce the short-hand notation $ \Theta(J_i)$ which denotes that $\theta_{J_i,J}$ is the largest angle of the $J_i$ relative to the collinear branch. In the first line, when $J_2$ fails, we proceed to test $J_1$ against the soft drop criterion (first term) while if $J_2$ passes, we stop (second term). In addition, we sum over all possible permutations. In the case of two branches this includes both configurations when $\theta_{J_2,J}$ and $\theta_{J_1,J}$ is larger. The second term in the second line of eq.~(\ref{eq:M2}) vanishes due to the contradiction of the two conditions $\theta_{J_2,J} > \theta_{J_1,J}$ and $\theta_{J_2,J} < R_g$ as required by ${\cal M}_1(J_2p)$ and $\theta_{J_1,J} > R_g$. 
We note that the angular ordering of the C/A algorithm is crucial here to generate the conflict. For the anti-k$_T$ algorithm, the overall $\Theta(J_i)$ is replaced by the anti-k$_T$ distance metric which reduces to $\Theta(J_i)$ up to power corrections since $\min(p_{T,J_i}^{-2\alpha},p_{T,J}^{-2\alpha}) \theta_{J_i,J} \sim p_{T,J}^{-2\alpha} \theta_{J_i,J}$ and the contradiction is still obtained. When other jet algorithms are used, the $\Theta$ here will be replaced by a different ordering, and the conflicts could therefore be avoided which would lead to a non-vanishing second term. This would eventually cause a difference between the jet veto and the soft drop declustering procedure. After carrying out the sum over the two permutations, we obtain the third line which is is an independent veto of the branches 1 and 2 when their separation from the collinear branch is larger than $R_g$. Thus the equivalence holds for 2 collinear-soft branches. Similarly, for 3 branches we have
\bea
{\cal M}_3 = &\sum_{\text{perm.}}\, \Theta(J_3) \left[ {\cal M}_2 \, {\cal M}_1(J_3  f)
+ {\cal M}_1(J_3 p)  \right]  \nn \\
=& \sum_{\text{perm.}}\,  \Theta(J_3)\left[ {\cal M}_2  \,  {\cal M}_1(J_3) \, 
+ \, {\cal M}_1( J_3 p ) (1 -   {\cal M}_2 ) \right] \nn \\
=& \, {\cal M}_1(J_1)  \,  {\cal M}_1(J_2)   {\cal M}_1(J_3) \,.
\eea
The first term in the first line states that if branch-$3$ fails the soft drop criterion, we proceed to test the remaining 2 branches until the procedure stops. The second term corresponds to the case where branch-$3$ passes the criterion and the algorithm terminates. Following a similar argument
as in the case of 2 branches, the second term in the second line vanishes and we get the last line which demonstrates the equivalence for $3$ branches. For arbitrary $N$, we find by induction that
\bea
{\cal M}_N  = 
  \prod_i^N {\cal M}_1(J_i)   \,.
\eea
This shows the equivalence between soft drop declustering and the jet veto procedure as long as $\theta_g \ll 1$ with small $z_{\rm cut}\ll 1$.

\subsection{The collinear function~\label{sec:collinearfunction}}

The operator definition of the collinear function as it appears in eq.~(\ref{eq:refactorize2}) can be written as
\bea
 \frac{\nslash}{2} C_{q,m} (\theta_g p_T R,\mu)  = &  \sum_{\rm spins} \prod_{j=1}^m
  \int \frac{d E_j E_j^{d-3}}{(2\pi)^{d-2}} | P_j (\{k_{X_c}\}) \rangle \langle P_j (\{k_{X_c}\}) |  \nn \\
& \times \,  2(2\pi)^{d-1} 
\,\delta(2E_J - {\bar n}\cdot k_{X_c} )\,\delta^{(d-2)} (k_{X_c}^\perp)\,\Theta\left( R_g  - \hat{r}_g|_{\rm C/A} \right) \,,
\eea
for quark jets and a similar expression can be obtained for gluon jets, see~\cite{Becher:2015hka}. Here the null four-vector is taken as ${\bar n} = (1, \hat{n})$, where $\hat{n}$ is pointing in the jet direction and $ | P_j (\{k_{X_c}\}) \rangle \langle P_j (\{k_{X_c}\}) |  $ 
is the matrix for producing the collinear state $X_c$. The measurement $\Theta\left( R_g   - \hat{r}_g|_{\rm C/A} \right)$ represents the C/A jet algorithm which acts on the final collinear state $X_c$ requiring that the separation between the last two branches in the clustering history is less than $R_g$. At NLO, after performing the angular integration, the collinear function as it appears in eq.~(\ref{eq:NLL}) is found to be
\bea
\langle C_i(\theta_g p_T R,\mu)  \rangle = 1 + \frac{\alpha_s}{2\pi}\left[ C_i\,
 \frac{L^2}{2}
+ \, \gamma_i  L+  d_i \right] \,,  
\eea
where $C_i$ on the right-hand side corresponds to $C_{F,A}$ for quarks and gluons, respectively. The other constants are given by
\bea
d_q &=C_F \left(\f{13}{2}-\f{3\pi^2}{4}\right)\,, \qquad \gamma_q = \f{3C_F}{2}\,, \label{eq:const1} \\ 
d_g &= C_A \left(\f{67}{9}-\f{3\pi^2}{4}\right)- T_F n_f \f{23}{9}\,, \quad \gamma_g = \f{\beta_0}{2} \,, \label{eq:const2}
\eea
and the logarithm $L$ is defined as
\bea
 L = \ln \left( \frac{\mu^2}{\theta_g^2\, p_T^2 R^2 }\right) \,,
\eea
see also~\cite{Ellis:2010rwa,Liu:2012sz}. The natural collinear scale choice used to minimize the logarithmic contribution is given by $\mu_C\sim \theta_g \, p_T R$, and the anomalous dimensions $\gamma_{C_i}$ are found to be
\bea
\gamma_{C_i}(\theta_g p_T R,\mu) = \frac{\alpha_s}{\pi} \left[ \gamma_i + C_i L\right] \,.
\eea

\subsection{The collinear-soft function~\label{sec:collinearsoftfunction}}

The collinear-soft function as it appears in eq.~(\ref{eq:refactorize2}) is defined at the operator level as
\bea
S^{\in \text{gr}}_{i,m} (z_{\rm cut}\theta_g^{1+\beta}p_T R,\mu;\beta) = 
\sum_{X_{cs}}\, \Theta\left( R_g   - \hat{r}_g|_{\rm soft drop} \right)
\, 
\left| \langle 0| W_{\bar n}^\dagger W_{n_1}^\dagger \dots W_{n_m}^\dagger |X_{cs}\rangle
\right|^2  \,, 
\eea
with the null vector $n_i = (1,\hat{n}_i)$, where $\hat{n}_i$ is oriented along the propagation direction of the collinear radiation $i$ and $W_n$ is a Wilson line in the $n$-direction. Here $ \Theta\left( R_g  - \hat{r}_g|_{\rm soft drop} \right)$ encodes the soft drop grooming algorithm operating on
the collinear-soft final state $X_{cs}$ with the knowledge of the eikonal directions $n_1 \dots n_m$. In order to achieve the resummation at NLL accuracy, see eq.~(\ref{eq:NLL}), we need the collinear-soft function at NLO. After performing again the angular integration, we find
\begin{equation}
\langle S^{\in \text{gr}}_{i}(z_{\rm cut}\theta_g^{1+\beta}p_T R,\mu;\beta) \rangle 
= 1 - \f{\alpha_s C_i}{2\pi} \frac{1}{1+\beta} \left[\f{1}{2} \ln^2 \left(\f{\mu^2}{z_{\rm cut}^2 \theta_g^{2(1+\beta)} p_T^2 R^2}\right)
 - \f{\pi^2}{12}\right] \,.
\end{equation}
The natural scale of the collinear-soft mode is indeed found to be $\mu_S^{\rm gr} \sim z_{\rm cut} \theta_g^{1+\beta} p_T R$, and the anomalous dimensions are given by
\bea\label{eq:collinear_soft_anom_dim}
\gamma_{S^{\in \rm gr}_i} (z_{\rm cut}\theta_g^{1+\beta}p_T R,\mu;\beta) =  -\f{\alpha_s C_i}{\pi}\f{1}{1+\beta} \ln \left(\f{\mu^2}{z_{\rm cut}^2 \theta_g^{2(1+\beta)} p_T^2 R^2}\right)  \,.
\eea

\subsection{Leading NGLs including C/A clustering effects~\label{sec:ngl}}

The leading NGLs of the $\theta_g$ distribution can be readily inferred from the equivalence with the jet veto case. The leading NGLs originate from correlated strongly ordered emissions when the harder emission is inside the groomed jet while the softer one is outside and vetoed. We adopt the notation of~\cite{Dasgupta:2001sh}. At next-to-next-to-leading order (NNLO), when clustering effects due to the C/A algorithm are ignored, we thus have
\bea\label{eq:nglint}
{\cal S}_{i,\rm NGL}(L,\theta_g) =&1 -  C_i  C_A \left(\frac{\alpha_s}{2\pi}\right)^2\, \int\frac{dx_1}{x_1} \frac{dx_2}{x_2} \int_{1\in J} \mathrm{d} c_1\frac{  \mathrm{d} \phi_1}{2\pi}
\, \, \int_{2\notin J} \mathrm{d} c_2 \frac{ \mathrm{d} \phi_2}{2\pi} 
\, \, \nn \\ 
&\times\Theta(x_1-x_2) \, \Theta(x_2 - z_{\rm cut}\theta_g^\beta) \, \frac{ \cos \phi_2}{(1 - c_1 c_2 - s_1 s_2 \cos \phi_2 )\, s_{1} s_{2}}
\, \nn \\
&\approx1 - C_i C_A  \left(\frac{\alpha_s}{2\pi}\right)^2 \frac{\pi^2}{3} L^2 \,.
\eea
Where we introduced the notation $L = - \ln(z_{\rm cut} \theta^\beta_g)$ and the polar angles $c_i = \cos\theta_i$ and $s_i = \sin\theta_i$ of the two emissions at NNLO measured with respect to the groomed jet axis and their respective transverse momentum fractions relative to the total momentum of the jet $x_i=k_{Ti}/p_T$. Here, we also replaced the veto condition
\bea
 \Theta(x_2 - z_{\rm cut}(\theta_2/R)^\beta)~\to~\Theta(x_2 - z_{\rm cut}\theta_g^\beta)  \,,
\eea
which is valid for the leading NGLs. Comparing with the structure in eq.~(\ref{eq:NGLstructure}), we would obtain the first coefficient $S_{i,2}$ as 
\bea\label{eq:coef}
S_{i,2} = - C_i C_A \frac{\pi^2}{3}\,.
\eea
When clustering effects are included, this coefficient will be reduced since any soft emission outside $R_g$ that is clustered into  the groomed jet will not be subject to the veto condition. At NNLO, 
the clustering happens when the distance between the emissions inside and outside $R_g$ is smaller than the distance between the groomed jet axis and the radiation inside $R_g$. Therefore, to we need to insert the constraint
\bea
\Theta(d_{12} - d_1)   \,,
\eea
with 
\bea
 d_i = \eta_i^2 + \phi_i^2\,, \quad \quad
 d_{ij} = (\eta_i - \eta_j)^2 + (\phi_i - \phi_j)^2 
\,,
\eea
and the phase space which generates the NGLs will thus be reduced.
We thus have the following modified expression compared to eq.~\eqref{eq:coef} above
\bea\label{eq:nglint2}
S^{\rm C/A}_{i,2}(\theta_g) &=  - 4  C_i  C_A \,    \int_{1\in J} \mathrm{d} c_1   \frac{  \mathrm{d} \phi_1}{2\pi}
\int_{2\notin J} \mathrm{d} c_2 \frac{ \mathrm{d} \phi_2}{2\pi}
\, 
\frac{ \cos\phi_2}{(1 - c_1 c_2 - s_1 s_2 \cos\phi_2 )\, s_{1} s_{2}}
\, \Theta(d_{12} - d_1) 
\,.
\eea
In principle, the integral in eq.~(\ref{eq:nglint2}) can be evaluated numerically. Using the small angle approximation, we can approximate the distances $d_i$ and $d_{ij}$ as
\bea
d_i & = \eta_i^2 + \phi_i^2
\approx 2 \frac{ k_i \cdot p }{k_{Ti}\, p_{T} }
=  2 (1-c_i) \approx 
\hat{\theta}_i^2 R_g^2 \\
d_{ij} & = (\eta_i - \eta_j)^2 + (\phi_i - \phi_j)^2 
\approx 2 \frac{ k_i \cdot k_j }{k_{Ti} \, k_{Tj} } \nn \\ 
& = 2  (1-c_ic_j-s_is_j\cos\phi_2 ) \nn \\
& \approx 
(\hat{\theta}_1^2  + \hat{\theta}_2^2
-2 \hat{\theta}_1\hat{\theta}_2 \cos\phi_2) R_g^2 \,.
\eea
The relevant integral can then be approximated as
\bea\label{eq:nglsca}
S^{\rm C/A}_{i,2}(\theta_g) & \approx  - 4  C_i  C_A \,    \int_{0}^1 \mathrm{d} \hat{\theta}_1   \, 
\int_1^{1/\theta_g} \mathrm{d} \hat{\theta}_2  \int_0^{2\pi} \, \frac{ \mathrm{d} \phi_2}{2\pi}
\, 
\frac{ 2 \cos\phi_2}{\hat{ \theta}_1^2  + \hat{\theta}_2^2
-2 \hat{\theta}_1 \hat{\theta}_2 \cos\phi_2 }
\, \Theta( \hat{\theta}_2 - 2 \hat{\theta}_1  \cos\phi_2 ) 
\,, 
\eea
where we introduced the variable $\hat{\theta}_i = \theta_i \cosh (\eta_J)/R_g$. Note that the variable change removes the dependence on $\eta_J$. Following the definition of the collinear-soft mode, the upper bound for the $\hat{\theta}_2$ integral should be $\infty$. Here we use instead $1/\theta_g=R/R_g$ keeping in mind that the radiation outside the groomed jet is within the original ungroomed jet with radius $R$, see Fig.~\ref{fig:correlation}. In addition, in the limit $\theta_g \to 1$ the associated NGLs in $\mathcal{F}_i$ vanish, as the NGLs are proportional to the area of the veto region. The integral in eq.~(\ref{eq:nglsca}) can be performed analytically for $\theta_g \ll 1$ and we find that the coefficient of the non-global logarithm is significantly reduced due to the additional constraint. In the limit $\theta_g\ll 1$, the $\theta_g$ dependence of $S_{i,2}^{\rm C/A}$ is power suppressed and we find 
\be\label{eq:clustering49}
S_{i,2}^{\rm C/A}(\theta_g) = -C_i C_A \f{\pi^2}{3}  \times \frac{4}{9} + \mathcal{O}(\theta_g)\,.
\ee
Thus the size of the NGL is reduced due to clustering by a factor of $4/9$. A similar reduction due to clustering effects was observed for example in~\cite{Delenda:2012mm} in the context of jet mass measurements. For general $R_g < R$, we find that the coefficient in eq.~(\ref{eq:nglsca}) evaluates to
\begin{align}
\label{eq:gen_Rg_R_fixed_two_loop_NGL}
 S^{\rm C/A}_{i,2}(\theta_g)
 =&  -4 C_i C_A \Bigg[ \frac{4}{9} \frac{\pi^2}{12}
+F(\theta_g) \, \Theta\Big(\theta_g-\f12\Big)  + \ln (1-\theta_g) \ln(\theta_g)  \nn \\
&+ {\rm Li}_2\left( - \frac{1}{\theta_g} \right) - {\rm Li}_2\left( - \frac{ 1 - \theta_g   }{\theta_g} \right)\Bigg]\,,
\end{align}
with 
\bea
F(\theta_g) = 
\int_{1/\theta_g}^2
\mathrm{d} y \int_0^{\cos^{-1} \!\! y/2}  \, 
\frac{\mathrm{d} \phi }{\pi} \, 
 \frac{ \ln(\theta^2_g y^2 ) c_\phi  }{ 1+y^2 -2y c_{\phi} } \,,
\eea 
which reduces to eq.~(\ref{eq:clustering49}) in the limit $\theta_g\ll 1$. The $\phi$ integral here can be done analytically but the result is rather lengthy. The remaining $y$ integral can be evaluated numerically.

\subsection{Leading Abelian C/A clustering logarithms~\label{sec:ab}}

To study the Abelian C/A clustering effects, we start with two independent collinear-soft emissions with momenta $k_{1,2}$. In order to extract the leading Abelian logarithm, it suffices to consider the strongly ordered limit in which $p_T \gg k_{1} \gg k_{2}$ or $p_T \gg k_{2} \gg k_{1}$. The leading logarithms come from the configuration where the harder gluon is initially inside the groomed jet and the softer gluon is outside but within the initial ungroomed jet. The C/A clustering pulls the softer gluon into the jet and generates a mismatch with the real-virtual correction. In the small $\theta_g$ approximation, the NNLO contribution to the Abelian clustering reads
\bea\label{eq:LCL}
\mathcal{A}_{i,{\rm Abel.}}^{\rm{C/A}} (L,\theta_g) & =
1+ \left(\frac{ \alpha_s }{\pi} C_i \right)^2  \, 
 \, 
 \frac{1}{2!}
 \int_{z_{\rm cut} \theta_g^\beta}^{1}
\frac{  \mathrm{d} x_1}{ x_1} 
\frac{  \mathrm{d} x_2 }{ x_2}\,  \int_0^{1/\theta_g}
  \,\frac{ \mathrm{d}\hat{\theta}_1}{  \hat{\theta}_1 } 
 \,\frac{  \mathrm{d}\hat{\theta}_2 }{ \hat{\theta}_2 } 
 \int_0^{2\pi}
 \frac{ \mathrm{d}\phi_1}{\pi}
 \frac{ \mathrm{d}\phi_2}{\pi}
\, \Theta_{\rm C/A}  
\,,
\eea
where we use the same notation conventions as in eq.~(\ref{eq:nglint}) above. Terms that are power suppressed by $\theta_g$ are omitted and can be found in~\cite{Delenda:2012mm}. We have
\bea
\Theta_{\rm C/A}   &=  \Theta( d_1 - R_g^2)\, \Theta(R_g^2 - d_2) \,\Theta(d_{2 } - d_{12}) \, \nn \\
& \approx 
\Theta(\hat{\theta}_1^2  -1)\,
\Theta(1 - \hat{\theta}^2_2)\,
\Theta(  - \hat{\theta}_1^2 + 2 \hat{\theta}_1 \hat{\theta}_2 \cos\phi_2)  \,, 
\eea
where as before $\hat{\theta}_i = \theta_i \cosh (\eta_J)/R_g$. Performing the integral, we find the first Abelian corrections due to the C/A clustering
\begin{equation}\label{eq:Clustering_Final_result}
A_{i,2}^{\rm{C/A}} (\theta_g) = 
\frac{4C_i^2}{2!} \bigg[ 
\frac{\pi^2}{54}  - 2\,  \Theta\Big(\theta_g - \frac{1}{2} \Big) \int_0^{\cos^{-1}\frac{1}{2\theta_g}} \frac{ \mathrm{d} \phi}{\pi} \ln^2 ( 2 \theta_g c_{\phi})
 \bigg] \,.
\end{equation} 
The remaining integral can be worked out analytically but it is rather lengthy. In the small $\theta_g$ limit, we find
\bea
A_{i,2}^{\rm{C/A}} (\theta_g \ll 1) = \frac{4C_i^2}{2!} \frac{\pi^2}{54}\,.
\eea
Higher order coefficients $A_{i,3}^{\rm C/A}$ and beyond can be computed following the method developed in~\cite{Delenda:2012mm}. The authors further proposed an exponentiation of Abelian clustering logarithms by calculating higher order corrections and analyzing the structure of the perturbative series. In this work, we instead resum the Abelian clustering logarithms at LL and leading color using the Monte Carlo method discussed in section~\ref{sec:3}, where we compare to the fixed order result derived here.

\subsection{Subleading NGLs and clustering logarithms}

\begin{figure}[t]\centering
\includegraphics[width=3in]{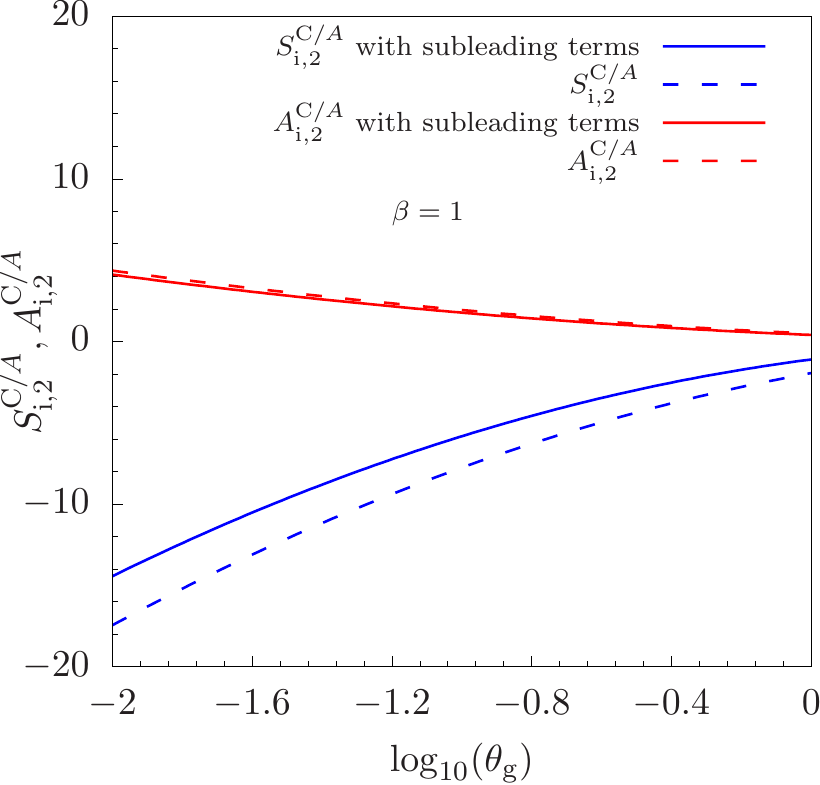} 
\caption{Comparison of the numerical size of the NGLs (blue) and the clustering logarithms (red) at NNLO with (solid) and without (dashed) subleading contributions as a function of $\theta_g$. The results are normalized to $(\alpha_s/\pi)^2 C_iC_A$ and $(\alpha_s/\pi)^2 C_i^2$ for the NGLs and the clustering logarithms, respectively. We choose the parameters $z_{\rm cut}=0.1$, $\beta=1$ as a representative example.~\label{fig:sub-leading}}
\end{figure}

Here we comment on the potential impact of subleading NGLs and Abelian clustering logarithms beyond single logarithmic accuracy. Due to the soft drop criterion, the corresponding jet veto parameter is $z_{\rm cut} d_i^{\beta/2}/R^\beta p_T 
= z_{\rm cut} \theta_g^{\beta} \hat{\theta}_i^\beta p_T$ with $\hat{\theta}_i$ as defined in the previous sections. Therefore, 
for instance the $x_i$-integration bound in eq.~(\ref{eq:LCL}) should be replaced by $z_{\rm cut} \theta_g^{\beta} \, \hat{\theta}_1^\beta$ instead of $z_{\rm cut}\theta_g^\beta$. The latter is sufficient to get the leading logarithms as derived in the previous section. The additional $\hat{\theta}_i$ dependence gives rise to part of the subleading logarithms. Similar reasonings apply to the NGL case. In the $\theta_g \to 0$ limit, we find at NNLO
\bea\label{eq:subngl}
{\cal S}_{i,{\rm NGL}}^{\rm C/A, sub} (L,\theta_g)= 
-4C_iC_A\left( \frac{\alpha_s}{2\pi} \right)^2 
\left(- 0.474 \, \beta \, L 
+ 1.0145 \frac{\beta^2}{4}\right)\,,
\eea
for the NGLs and 
\bea\label{eq:subclu}
{\cal A}_{i,{\rm Abel.}}^{\rm C/A, sub} (L,\theta_g)= 
\frac{1}{2} \left(\frac{\alpha_s C_i}{\pi} \right)^2
\left(- 0.073 \, \beta \, L 
+  0.045 \frac{\beta^2}{4} \right)\,,
\eea
for the Abelian clustering logarithms. In Fig.~\ref{fig:sub-leading}, we show a comparison of the leading NGL (blue) and clustering logarithms (red) at NNLO (dashed) and when the subleading terms in eqs.~(\ref{eq:subngl}) and~(\ref{eq:subclu}) are included (solid). We result is plotted as a function of $\theta_g$, for exemplary values of the soft drop grooming parameters $\beta=1,\; z_{\rm cut}=0.1$. We observe a moderate reduction of the NGL contribution when the subleading terms in eq.~(\ref{eq:subngl}) are included. The subleading contribution of the Abelian clustering logarithms turns out to be almost negligible.

Though not yet conclusive, the results in this section suggest that the impact of subleading NGLs and clustering logarithms is moderate. We thus expect that the numerical results for the soft drop groomed jet radius presented in section~\ref{sec:4}, which only include the leading NGLs and Abelian clustering logarithms to all orders, capture the dominant perturbative effects.

\subsection{Comparison to results in the literature}

In this section, we compare the calculation presented in this work to results available in the literature. In~\cite{Larkoski:2014wba}, the resummation of the cumulative $\theta_g$ distribution was realized at MLL accuracy. We show that our results reduce to~\cite{Larkoski:2014wba} when only the leading logarithms are taken into account. The resummation in~\cite{Larkoski:2014wba} is based on the coherent branching formalism and the result can be expressed as
\bea\label{eq:MLL}
\frac{1}{\sigma_{\rm incl}}\frac{{\rm d}\Sigma(\theta_g)}{{\rm d}p_T\, {\rm d}\eta} = f_q~\Sigma_q(\theta_g) + f_g~\Sigma_g(\theta_g) \,.
\eea
Here $\sigma_{\rm incl}$ denotes the inclusive jet cross section, $f_i$ are the leading-order quark/gluon fractions and $\Sigma_i(\theta_g)$ denote the respective resummed exponents that depend on $\theta_g$. On the right hand side we leave the dependence on other variables besides $\theta_g$ implicit. At MLL accuracy for a fixed coupling constant, the resummed exponent can be written as
\bea\label{eq:MLL_Sigma}
\Sigma_i(\theta_g) & \,\stackrel{{\rm f.c.}}{=}\, \exp\bigg[-\frac{\alpha_s}{\pi}C_i \bigg(\beta \ln^2\theta_g+2\ln z_{\rm cut}\ln\theta_g + \frac{\gamma_i}{C_i} \ln\theta_g\bigg)\bigg]\,.
\eea
where the constants $\gamma_i$ are defined in eqs.~(\ref{eq:const1}) and~(\ref{eq:const2}). At MLL accuracy also running coupling effects are taken into account. 

The improvements achieved in this work concern both the quark/gluon fractions $f_i$ and the resummed exponents $\Sigma_i(\theta_g)$. Here, the resummation is carried out at full NLL accuracy, including both global and non-global logarithms. In addition, clustering effects due to the C/A algorithm are taken into account. In order to recast the formalism developed here into the form of eq.~(\ref{eq:MLL}), we separate the production of the jet from the jet substructure measurement as discussed in~\cite{Banfi:2006hf,Kaufmann:2015hma,Cal:2019hjc}. We start by rewriting the jet function ${\cal G}_c$ in eq.~(\ref{eq:inclusive}) at fixed order as
\bea\label{eq:separate}
{\cal G}_c(z,\theta_g,p_T R, \mu; z_{\rm cut}, \beta)\;=& \;\sum_d J_{cd}(z,p_T R,\mu) \nn\\
&\times \,\int {\rm d}z \left[{\cal G}_d(z,\theta_g,p_T R, \mu; z_{\rm cut}, \beta)-J^{(1)}_d(z,p_TR,\mu)\right] + \,{\cal O}(\alpha_s^2)\,.
\eea
Here $J_d^{(1)}$ is the ${\cal O}(\alpha_s)$ contribution of the semi-inclusive jet function as it appears in the inclusive jet cross section~\cite{Kaufmann:2015hma,Kang:2016mcy,Dai:2016hzf}. The functions $J_{cd}$ are related to the semi-inclusive jet functions except that we keep track also of the jet flavor $d$ such that
\be
\sum_d J_{cd}(z,p_T R,\mu)=J_c(z,p_T R,\mu) \,.
\ee
We would like to stress that only at leading-order the jet flavor $d$ is the same as the final state parton $c$ from the hard-scattering event. Note that the separation in eq.~(\ref{eq:separate}) is multiplicative and the functions $J_{cd}$ contain the complete $z$-dependence. The $z$-dependence is associated with out-of-jet radiation diagrams at NLO and it is the same for different jet substructure observables. We can now calculate the cross section for a jet of flavor $d$ as
\be\label{eq:flavorinclusive}
\f{{\rm d}\sigma_d}{{\rm d}\eta \, {\rm d}p_T} = \sum_{abc} f_a\otimes f_b\otimes H_{ab}^c\otimes J_{cd}\,.
\ee
Here we use a more compact notation compared to the factorization in eq.~(\ref{eq:inclusive}) above. After summing over $d$ in eq.~(\ref{eq:flavorinclusive}), the inclusive jet cross section $\sigma_{\rm incl}$ is obtained. Therefore, in our calculation 
the quark/gluon fractions $f_i$, see eq.~(\ref{eq:MLL}), can be obtained systematically beyond leading-order as
\be\label{eq:flavorfraction}
f_{q(g)} = \frac{1}{\sigma_{\rm incl}} \, \sum_{abc} f_a\otimes f_b\otimes H_{ab}^c\otimes J_{cq(g)}\,,
\ee
where also the $\ln R$ resummation is included. The resummed exponents beyond MLL accuracy are now obtained from the refactorized expression of the jet function ${\cal G}_{q,g}$ after subtracting the NLO semi-inclusive jet function at fixed order, see eq.~(\ref{eq:separate}). Following the discussion in the sections above, we thus have
\bea
\Sigma_i(\theta_g)\,&=\, \langle \tilde H_i(p_T R,\mu)\rangle \, \langle S^{\notin \text{gr}}_{i}(z_{\rm cut}p_T R,\mu;\beta)\rangle \, \langle C_{i}(\theta_g\, p_TR,\mu)\rangle \, \langle S^{\text{gr}}_{i}(z_{\rm cut} \theta_g^{1+\beta}\, p_T R,\mu;\beta) \rangle \, . \nn\\
\eea
The constants $\tilde H_i$ were calculated in~\cite{Cal:2019hjc}. After solving the evolution equations of the different functions and including NGLs and Abelian clustering logarithms, the resummation at NLL accuracy can be achieved which includes logarithms of $\theta_g$, $R$ and $z_{\rm cut}$. The result for fixed scales at leading logarithmic accuracy is given by
\begin{align}\label{eq:exponentscales}
   \langle \tilde H_i(p_T R,\mu)\rangle \exp\left[-\frac{\alpha_s C_i}{\pi} \left(\frac{1}{1+\beta}\left(\ln^2\frac{\mu_{\mathcal{H}}}{\mu_{S{\notin {\rm gr}}}}-\ln^2\frac{\mu_{\mathcal{H}}}{\mu_{S{\in {\rm gr}}}}\right)+\ln^2\frac{\mu_{\mathcal{H}}}{\mu_C}\right)+\frac{\alpha_s\gamma_i}{\pi}\ln\frac{\mu_{\mathcal{H}}}{\mu_C}\right] \,.
\end{align}
After making the canonical scale choices, which we list here for convenience
\bea
    \mu^{\rm can}_{\cal H}\,&=\,p_T R  \,,\label{eq:sc1}\\
    \mu^{\rm can}_{S{\notin {\rm gr}}}\,&=\, z_{\rm cut}\, p_T R  \,,\\
    \mu^{\rm can}_{C}\,&=\, \theta_g\,  p_T R  \,,\\
    \mu^{\rm can}_{S{\in {\rm gr}}}\,&=\, z_{\rm cut}\theta_g^{1+\beta} p_T R  \,,\label{eq:sc2}
\eea
we recover the result for $\Sigma_i(\theta_g)$ in eq.~(\ref{eq:MLL_Sigma}) up to the constants $\tilde H_i$, which is 1 at leading order. As can be seen from eq.~(\ref{eq:exponentscales}), the terms containing double logarithms of $z_{\rm cut}$ in the exponent of $\Sigma_i(\theta_g)$ can generally induce contributions to the QCD scale variations considered in the next section. Only for the central scale choice or when the scales $\mu_{S \notin {\rm gr}}$ and $\mu_{S\in {\rm gr}}$ are varied simultaneously, the contribution of these logarithms cancel completely. 
Of course we further include NGLs and Abelian clustering logarithms to achieve full NLL. Furthermore, we would like to stress again that an important feature of our approach is that it can be systematically extended beyond NLL accuracy. 

\section{The soft drop groomed radius in Monte Carlo~\label{sec:3}}

In this section, we present an algorithm for the large-$N_c$ leading log resummation of the NGL distribution, including clustering effects. We also perform a numerical comparison to the resummed distributions to gauge power corrections in factorizing the NGLs of the soft function of eq.~\eqref{eq:refactorize} from those of the collinear-soft function of eq.~\eqref{eq:refactorize2}, as well as the range of validity approximating the all orders resummation with the two-loop leading NGLs and Abelian clustering logarithms calculated previously.

\subsection{The Monte Carlo setup}

As explained in section~\ref{sec:proof}, the soft drop declustering angle operates as a jet veto algorithm. However, the jet being vetoed is simply the last branch to be declustered in the C/A algorithm once that branch is at an angular scale larger than $R_g$. The ungroomed jet is defined by the anti-$k_t$ algorithm, so the jet has a hard angular boundary at $R$, whereas $R_g$ is the soft drop declustering angle. Then the Monte Carlo resummation in the large-$N_c$ limit follows the general procedure found in~\cite{Dasgupta:2001sh,Appleby:2002ke,Neill:2018yet}. We define:
\begin{itemize}
   \item $t$ as the MC time \begin{align}t=\frac{C_A}{2\pi}\int\displaylimits_{\omega}^{Q}\frac{{\rm d}\mu}{\mu}\alpha_s(\mu)\,,
   \end{align} 
   where $\omega$ is the energy of the emission established at the shower time $t$. In our case, the scale $Q$ is set by $p_T R$.
  \item A histogram $H_t$ indexed by $t$.
  \item $R_g$ as the subjet radius, $R$ as the fat jet radius.  
  \item $\mathcal{D}$ as the list of dipoles.
  \item $k$ is the current number of emissions.
  \item $n_P$ is the direction of the first branch in the declustering procedure that passes soft drop. $n\cdot n_P$ sets the current angular scale of the shower.   
  \item $\mathcal{E}_J$ is the list of emitted eikonal lines that cluster into either the jet direction $n$ or $n_P$.
\end{itemize}
All eikonal lines are of the form $n_i=(1,\hat{n}_i)$, so the Lorentz product $n_{i}\cdot n_{j}=1-\cos\,\theta_{ij}$ simply measures the angle between the lines. Strong energy ordering implies the following clustering rule, ignoring recoil:
\begin{align}\label{eq:cluster_rule}
\{\omega_i,n_i\}+\{\omega_j,n_j\}\rightarrow \begin{cases}\{\omega_i,n_i\}\text{ if } \omega_i>\omega_j\\ \{\omega_j,n_j\}\text{ if } \omega_j>\omega_i\end{cases}
\end{align}
where $\omega_{i,j}$ are the energies and $n_{i,j}$ are the null directions of the two emissions to be clustered. Since the shower is energy ordered, we only need to keep track of the order of emissions to know the relative energy scales. We initialize the shower as follows:
\begin{itemize}
\item $t=0$.
\item $\mathcal{E}_J=\{n\}$.
\item $\mathcal{D}=\Big\{\{n,\bar{n}\}\Big\}$ for a quark jet, $\mathcal{D}=\Big\{\{n,\bar{n}\},\{\bar{n},n\}\Big\}$ for a gluon jet.  
\end{itemize}  
The algorithm then proceeds as:
\begin{enumerate}
\item Take an MC time step $t\rightarrow t+\Delta t$ and create a new emission with energy and direction $\{\omega_{k+1},n_{k+1}\}$. For details, see App.~\ref{App:dipole_split}.
\item Check if $n_{k+1}\cdot n > 2\,\sin^2(R/2)$, if this is true, the emission is outside the jet, goto step 1.   
\item Check if $n_{k+1}\cdot n_j > n_P\cdot n,\, \forall n_j\in \mathcal{E}_J$.
  \begin{itemize}
  \item If at least one of these conditions fail, add $n_{k+1}$ to $\mathcal{E}_J$. Goto step 1.
  \item If all these conditions are true, the emission is a new candidate for the declustering branch.
    \begin{itemize}
      \item If $n_{k+1}\cdot n < 2\,\sin^2(R_g/2)$, set $n_P$ to $n_{k+1}$, add $n_{k+1}$ to $\mathcal{E}_J$. Goto step 1. 
      \item If $n_{k+1}\cdot n > 2\,\sin^2(R_g/2)$. Add 1 to appropriate bin of $H_t$, and re-initialize shower for a new event. 
    \end{itemize}  
  \end{itemize}  
\end{enumerate}
Finally, normalize the histogram by the number of events. 

\subsection{Vetoing the declustered branches}

\begin{figure}[t]\centering
\includegraphics[width=3.3in]{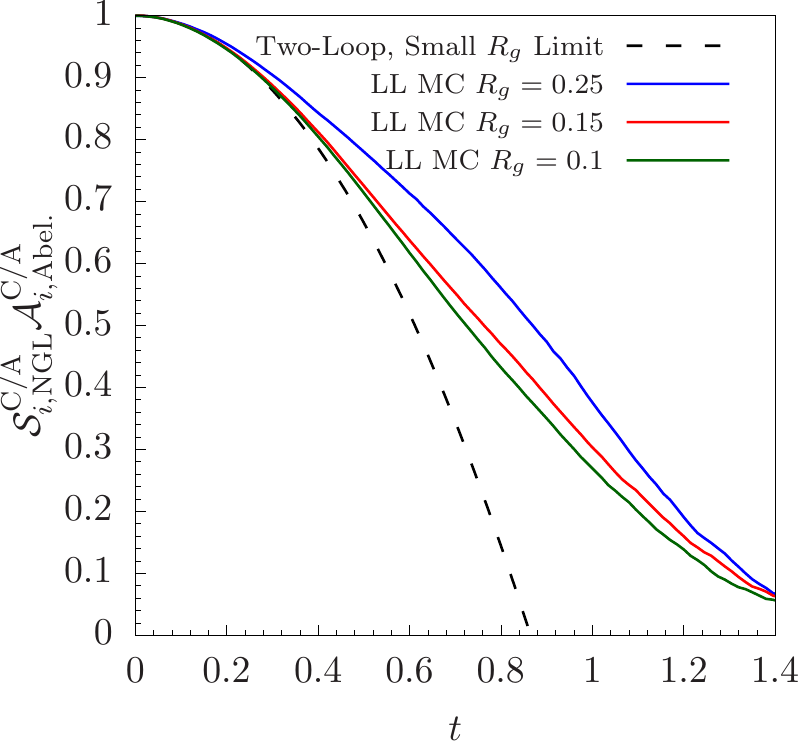} 
\caption{The NGL+clustering distribution at large-$N_c$ at LL for an initial fundamental (quark) dipole at various soft drop angles, compared to the small $R_g$ limit of the NNLO leading NGL of eq.~\eqref{eq:clustering49} and large-$N_c$ limit of the clustering effects in eq.~\eqref{eq:Clustering_Final_result}.~\label{fig:large_t_resum}}
\end{figure}

We construct $\mathcal{E}_J$ such that all lines within will cluster into either $n_P$ or $n$ before $n_P$ and $n$ themselves cluster at each step in the shower. Thus if we are given a new emission $n_{k+1}$ such that $n_{k+1}\cdot n_j > n_P\cdot n,\, \forall n_j\in \mathcal{E}_J$, then $n$ and $n_P$ will cluster before $n_{k+1}$ clusters into any of the established eikonal lines. Thus $n_{k+1}$ and $\mathcal{E}_J$ now define the two branches that are the first to be declustered under C/A, and the branch formed from $\mathcal{E}_J$ will have direction $n$ according to the clustering rule eq.~\eqref{eq:cluster_rule}. We then check whether the angle between these two branches is less than the desired $R_g$. If it is, we redefine the branch $n_P$ to be $n_{k+1}$, this is the new branch that sets the current declustering angle. If $n_{k+1}$ is at too wide an angle from $n$, then the emission $n_{k+1}$ sets the energy scale $z_{\rm cut} p_T R_g $. If we were to create subsequent emissions in the shower, they would have energy below $z_{\rm cut} p_T R_g $, and so if they created new branches, they would fail soft drop, and if they are clustered into the branches which pass soft drop, they cannot change the directions of those branches according to the clustering rule eq.~\eqref{eq:cluster_rule}. Thus $n_P$ cannot change, and the shower is over.

If $n_{k+1}\cdot n_j < n_P\cdot n,\,$ for at least one $n_j\in \mathcal{E}_J$, this emission in the shower will cluster into an emission that will eventually cluster into either $n$ or $n_P$ \emph{before} $n$ and $n_p$ themselves cluster. Thus this emission will not change the direction of the two branches that pass soft drop.

\subsection{Numerical results}

\begin{figure}[t]\centering
\includegraphics[width=3.3in]{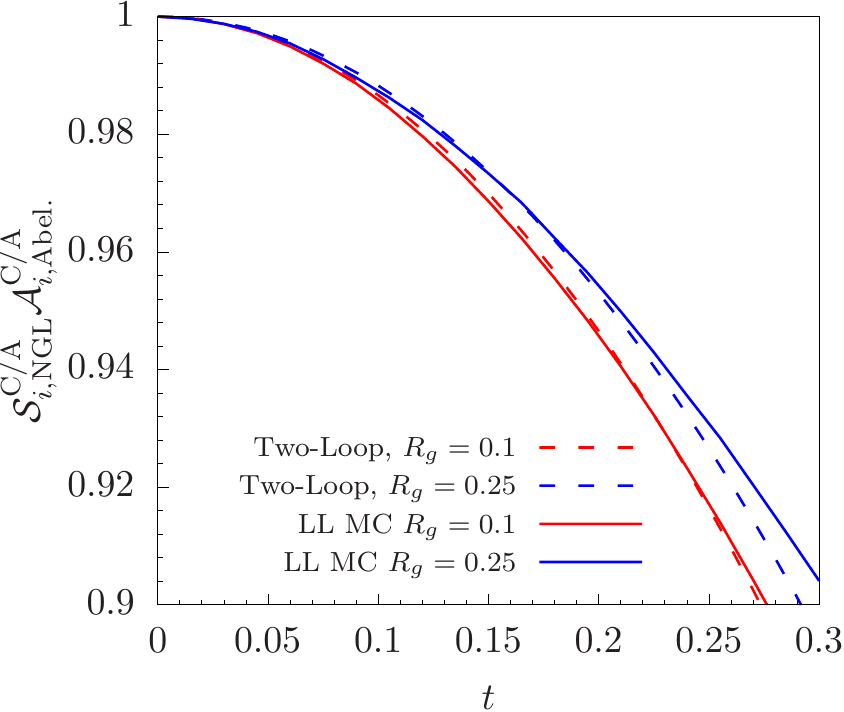} 
\caption{The NGL+clustering distribution at large-$N_c$ at LL for an initial fundamental (quark) dipole, comparing the two-loop results of eq.~\eqref{eq:gen_Rg_R_fixed_two_loop_NGL} and eq.~\eqref{eq:Clustering_Final_result}, with $C_F\rightarrow C_A/2$, to the all orders resummation at groomed angles $R_g=0.25$ and $0.1$.~\label{fig:small_t_resum}}
\end{figure}

 Formally, the Monte Carlo algorithm described above resums the NGLs from both the collinear-soft function defined in eq.~\eqref{eq:refactorize2}, and the soft function of eq.~\eqref{eq:refactorize}. However, in the small $R_g$ limit, these two functions factorize from each other. Thus to isolate the NGLs from the collinear-soft function alone, we divide out from the histogram produced by the LL MC described above both the hemisphere jet-mass NGL distribution of Ref. \cite{Dasgupta:2001sh} (which corresponds to the NGLs of the soft function of eq.~\eqref{eq:refactorize}), as well as dividing out the exponentiation of the one-emission contribution to the distribution to remove any global contributions. These one-emission contributions are included in the anomalous dimension calculated in eq.~\eqref{eq:collinear_soft_anom_dim}. For an initial quark dipole the distributions for $R_g=0.25,0.15,0.1,0.05$ are shown in Fig. \ref{fig:large_t_resum}, with the ungroomed jet radius of $R=0.8$, and the small-$t$ region is highlighted in Fig. \ref{fig:small_t_resum}. We have check numerically that the gluon distribution with an adjoint dipole is well approximated by the square of the quark distribution, despite clustering effects which would spoil this relation at large-$N_c$. In comparison to the fixed order results, we include in the large-$N_c$ ($C_F\rightarrow C_A/2$) limit both the contributions from eq.~\eqref{eq:gen_Rg_R_fixed_two_loop_NGL} and eq.~\eqref{eq:Clustering_Final_result}, since the MC covers the whole soft phase space at large-$N_c$ and leading log. Thus the MC algorithm accounts for clustering effects off the primary emission, but only in the large-$N_c$ limit. Using the methods of~\cite{Banfi:2005gj}, we could resum the Abelian clustering effects with the correct color structure, accounting for some of the subleading $N_c$ effects.
 
 Since the MC includes the multiple emissions evolution in the out-of-jet region as well as the evolution in the groomed region, we can test this collinear factorization of the two soft functions. We can see that for multiple emissions at $R_g=0.25, R=0.8$ the power corrections to the small $R_g$ limit of the collinear function are sizeable. However, for $R_g \leq 0.15, R=0.8$, the small $R_g$ limit of the fixed order NGL at two-loops distribution describes well the NGL distribution for phenomenological values of $t$. Moreover, we have checked that the LL distribution for the collinear-soft function is independent of $R$ once we are in the regime $R_g\ll R$. For example, the distribution for $R_g=0.1$ and $R=1.5708$ is almost identical up to statistical noise as $R_g=0.1$ and $R=0.8$. We use a shower angular cutoff scale of $\delta = 0.001$, and checked the independence of the distributions.

\subsection{Evolving dipoles~\label{App:dipole_split}}

We start with a list of dipoles $\mathcal{D}$, where an element is given by $\{x,y\}$. $x,y$ are the null directions forming eikonal lines of the dipole. We let:
\begin{align}
W_{x y}^\delta(j)&=\Theta\left(\theta_{xj}-\delta\right)\Theta\left(\theta_{yj}-\delta\right)\frac{x\cdot y}{(x\cdot j)\,(j\cdot y)}\,,\\\nonumber\\
P_{xy}^{\delta}&=\int\frac{d\Omega_j}{4\pi}W_{xy}^{\delta}(j)\approx\text{ln}\bigg(4\frac{\sin^2\frac{\theta_{xy}}{2}}{\delta^2}\bigg)+O(\delta^2)\,.
\end{align}
Then
{\small\begin{enumerate}
\item Calculate $P_{\mathcal{D}}^{\delta}$ by summing over the $P_{xy}^{\delta}$'s calculated from each dipole in $\mathcal{D}$. Uniformly generate a random number $\text{{\bf rnd}}\in[0,1]$, and then $\Delta t$ is determined by
\begin{align}
\Delta t&=-\frac{\text{ln} (\text{{\bf rnd}})}{C_A P_{\mathcal{D}}^{\delta}} \,.
\end{align}
Increase $t$ by $\Delta t$.
\item Choose the dipole $\{x,y\}\in \mathcal{D}$ with probability $P_{xy}^{\delta}/P_{\mathcal{D}}^{\delta}$.
\item Create an emission $j$ with distribution $W_{xy}^{\delta}(j)$, such that the angles are $\theta_{xj},\theta_{yj}>\delta$. This is equivalent to uniformily randomly creating in the rest frame of $x,y$ a direction $j$ in rapidity and azimuth, then boosting back to the lab frame and checking the angular constraint.
\item To split the dipoles again: delete $\{x,y\}$ from $\mathcal{D}$, add $\{x,j\}$ and $\{j,y\}$ to $\mathcal{D}$, then repeat splitting process according to phase space constraints.  
\end{enumerate}}

\section{Phenomenology~\label{sec:4}}

\bef
\includegraphics[width=5.7in]{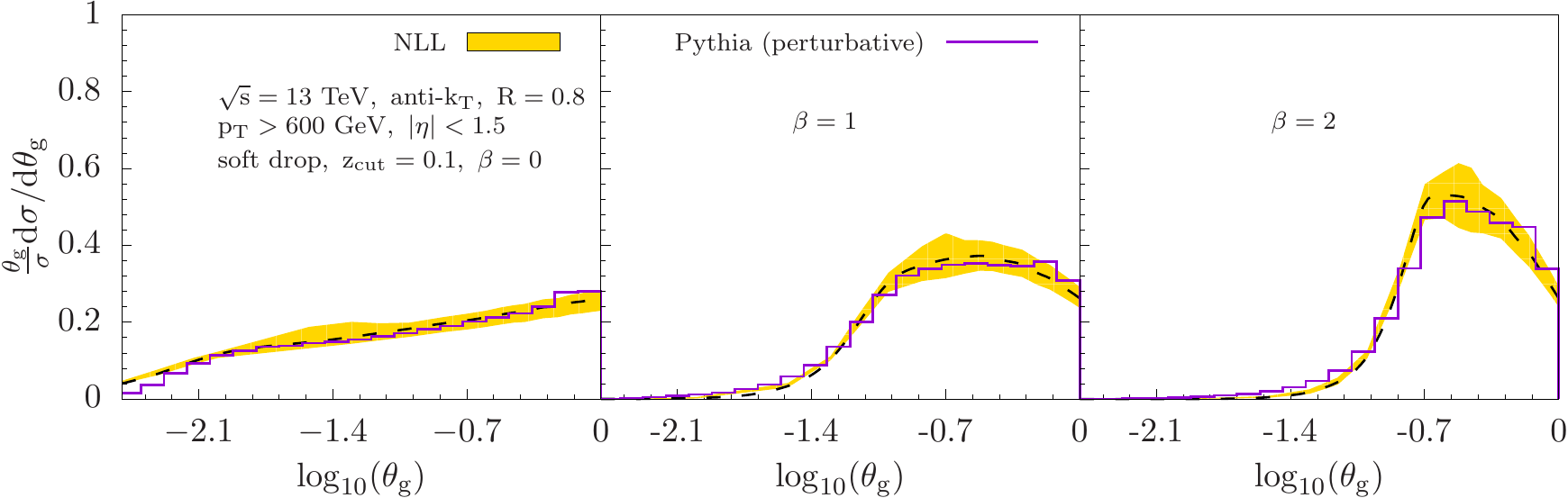} 
\caption{The soft drop groomed jet radius $\theta_g$ at NLL (dashed black, yellow band) in proton-proton collisions at $\sqrt{s}=13$~TeV. The jet kinematics are specified in the figure. We compare to Pythia 8 simulations~\cite{Sjostrand:2014zea} (purple) at the parton level, without hadronization and the underlying event contribution, for three different values of $\beta=0,\,1,\, 2$ (left to right).~\label{fig:parton}}
\eef

In this section we present numerical results for the soft drop groomed jet radius $\theta_g=R_g/R$ at NLL accuracy. We start with proton-proton collisions at $\sqrt{s} = 13\, {\rm TeV}$ collision and we consider inclusive jets $pp\to{\rm jet}+X$ with transverse momentum $p_T>600$~GeV in the central rapidity region of $|\eta|<1.5$. Jets are reconstructed with the anti-$k_T$ algorithm and $R=0.8$. For all numerical results presented in this section we choose the soft threshold parameter $z_{\rm cut} = 0.1$ and we present results for different values of the angular exponent $\beta$. We choose the NLO CT14 PDF set of~\cite{Dulat:2015mca} as default for all our numerical calculations. Since we consider the $\theta_g$ differential cross section normalized to the inclusive jet production cross section, the impact of the choice of the PDF set is small. In Fig.~\ref{fig:parton}, we present the numerical results (dashed black, yellow band) for the $\theta_g$ differential distribution $\theta_g/\sigma_{\rm incl}\, {\rm d}\sigma/{\rm d}\theta_g$ which is obtained by differentiating the cumulative cross section $\Sigma(\theta_g)$, see eq.~(\ref{eq:diff}). Before taking the derivative with respect to $\theta_g$, we choose the canonical scales as listed in eqs.~(\ref{eq:sc1})-(\ref{eq:sc2}) and we evolve all relevant functions that appear in the refactorization theorem to a common scale. The three panels show the result for different values of $ \beta= 0,\, 1,\,2$ (left to right). The QCD scale uncertainties as shown by the yellow band in Fig.~\ref{fig:parton} are obtained by varying all scales by factors of 2 around their canonical scales while maintaining the relations
\bea
\frac{1}{2} \leq \frac{\mu_i}{\mu_i^{\rm can}}\big/\frac{\mu_j}{\mu_j^{\rm can}}\leq 2\,
\eea
and
\bea
\mu^{\rm}_{S{\notin {\rm gr}}} &= z_{\rm cut}\, \mu^{\rm}_{\cal H}  \,,\\
\mu^{\rm}_{S{\in {\rm gr}}}&= z_{\rm cut}\, \theta_g^\beta \, \mu^{\rm}_C  \,.
\eea
\bef
\includegraphics[width=5.7in]{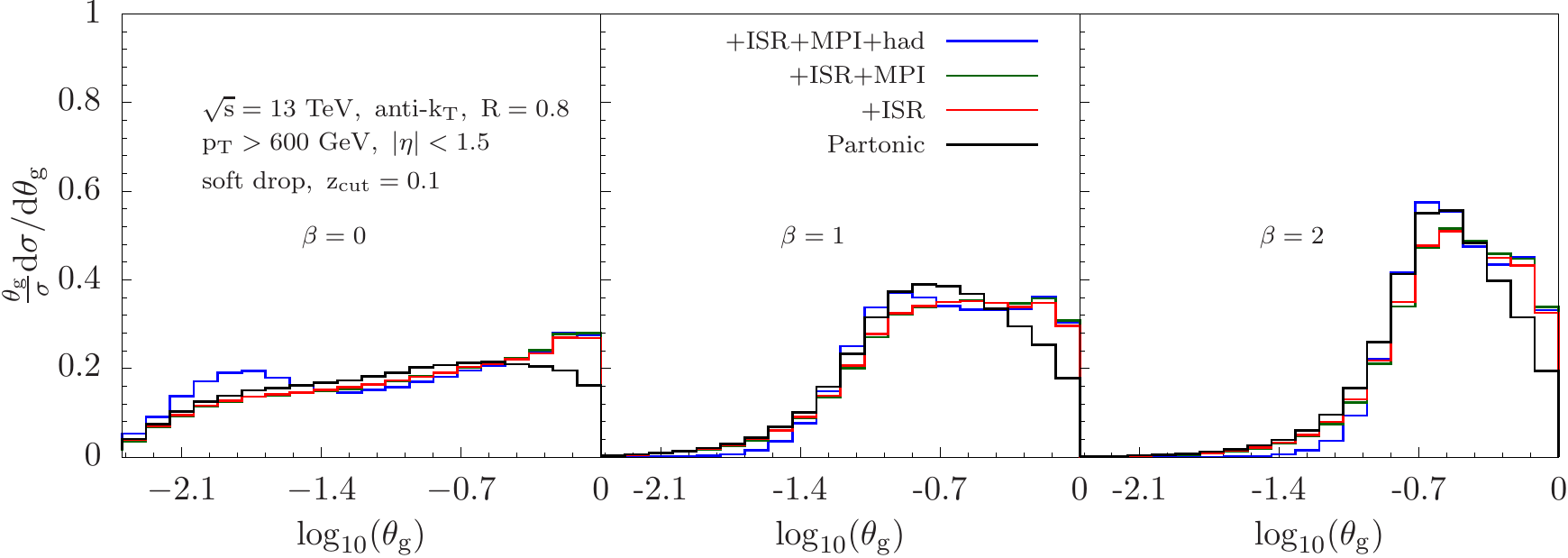} 
\caption{Pythia 8 results~\cite{Sjostrand:2014zea} for the soft drop groomed jet radius $\theta_g$ for the same kinematics as in Fig.~\ref{fig:parton} above. We separately show the purely partonic result (black), including initial-state radiation (red), multi-parton interactions (green) and hadronization corrections (blue).~\label{fig:pythia}}
\eef
As expected, we find that aggressive soft drop grooming ($\beta=0$) yields a relatively flat distribution (multiplied by $\theta_g$) of the soft drop groomed jet radius. Less aggressive grooming ($\beta=1,\,2$) instead gives a distribution that peaks at intermediate to relatively large values of $\theta_g$ which means that the groomed jet does not shrink as much in size compared to the initial ungroomed jet. Eventually, in the limit $\beta\to\infty$, the groomer is removed and the distribution approaches a delta function at $\theta_g=1$. For comparison, we also show Pythia 8.230 results using the default tune~\cite{Sjostrand:2014zea} (purple) in Fig.~\ref{fig:parton}. Here we do not include the nonperturbative hadronization and the contribution from underlying event (UE). Below we study these effects in more detail. In general, we find very good agreement between the Pythia 8 simulation and our perturbative results at NLL accuracy concerning both the shape and the overall magnitude. We note that in the region $\theta_g\sim 1$ perturbative power corrections play a role which is not captured by our factorization theorem. In principle, they could be included at fixed order by performing a matching calculation. In addition, we note that the perturbative resummation region ends when the lowest scale in the factorization theorem runs into the nonperturbative regime $\mu_{S\in {\rm gr}}\sim z_{\rm cut}\theta_g^{1+\beta}p_T R\sim\Lambda_{\rm QCD}\sim 1$~GeV. This corresponds to values of the soft drop groomed jet radius of $\theta_g\lesssim (\Lambda_{\rm QCD}/(z_{\rm cut}p_T R))^{1/(1+\beta)}$. Numerical results in this region are obtained by smoothly freezing the running of the QCD coupling constant above the Landau pole. We choose to freeze the running of $\alpha_s$ at the scale $0.4$~GeV for the numerical results presented here.

In Fig.~\ref{fig:pythia}, we study different perturbative and nonperturbative effects for the same kinematics as in Fig.~\ref{fig:parton}. First, the purely partonic results are shown by the black line. We then include the contribution of Initial State Radiation (ISR) as shown by the blue line. In general, ISR leads to a shift of the distribution toward larger values of $\theta_g$. We note that there is no exact one-to-one correspondence between the different parton and hadron level contributions in Pythia and the QCD factorization theorem we develop in this work.  ISR is power suppressed and not included in our factorization theorem. However, since ISR is a perturbative initial state contribution, we include it in the comparison in Fig.~\ref{fig:parton} above. Next, we include the underlying event contribution or multi-parton interactions (MPI), as shown by the green line in Fig.~\ref{fig:pythia}. As it turns out, MPI does not play a significant role for the jet kinematics and soft drop parameters considered here. Finally, we include hadronization effects for the $\theta_g$ distribution as shown by the blue line in Fig.~\ref{fig:pythia}. Interestingly, hadronization leads to a shift of the distribution to lower values of $\theta_g$ for soft drop grooming with $\beta=0$, whereas it shifts the distribution to higher values for $\beta=1,\, 2$. It will be interesting to study such effects in more detail in the future following the work of~\cite{Hoang:2019ceu}. We conclude that the impact of different effects that are not captured by the factorization theorem presented in this work are relatively small and that the soft drop groomed radius $\theta_g$ is under good control within perturbative QCD.

\bef
\includegraphics[width=5.7in]{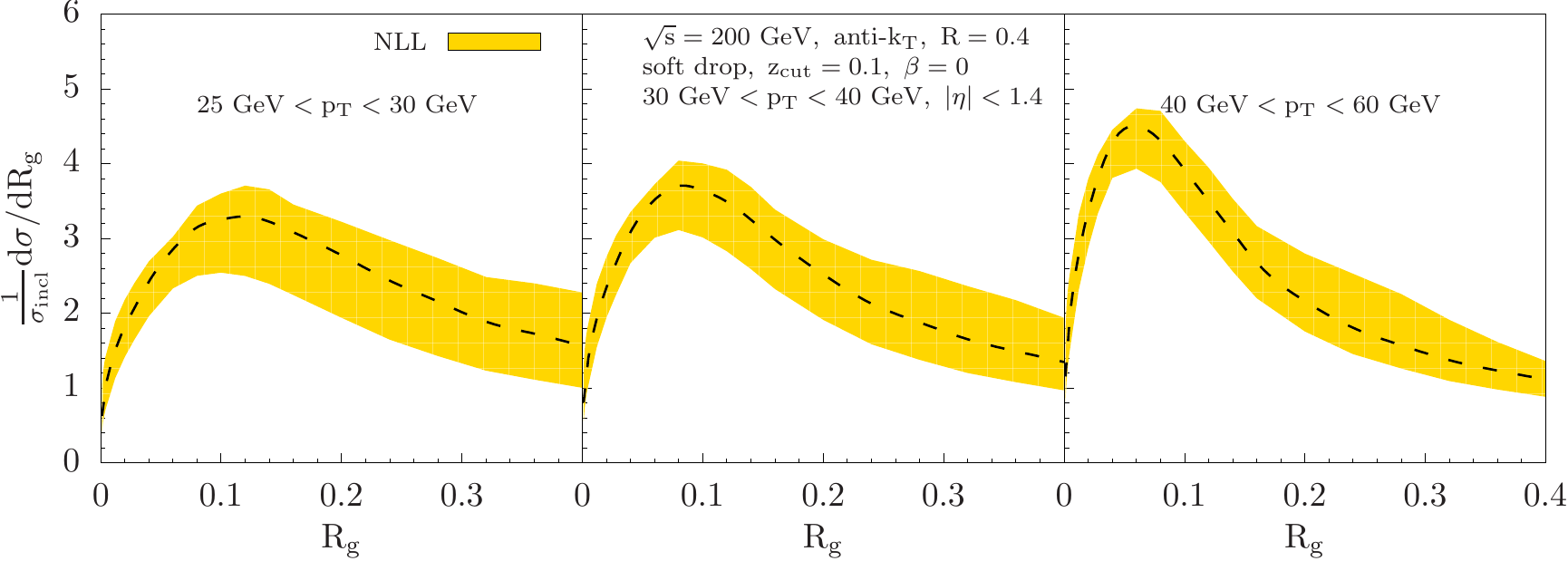} 
\caption{The soft drop groomed jet radius $R_g$ at NLL in proton-proton collisions for STAR kinematics at $\sqrt{s}=200$~GeV. The jet kinematics and soft drop parameters are indicated in the figure.~\label{fig:star}}
\eef

Finally, in Fig.~\ref{fig:star} we present results for jet kinematics relevant for the STAR experiment at RHIC at $\sqrt{s}=200$~GeV. For the setup of the analysis presented in~\cite{KunnawalkamElayavalli:2019wrv}, we show numerical results for three jet transverse momentum intervals $25<p_T<30$~GeV, $30<p_T<40$~GeV and $40<p_T<60$~GeV (left to right) with $R=0.4$, $z_{\rm cut}=0.1$ and $\beta=0$. We show the theoretical result $1/\sigma_{\rm incl}{\rm d}\sigma/{\rm d}R_g$ as a function of $R_g$. As expected, the QCD scale uncertainty is relatively large for the lower jet transverse momentum intervals considered here.
 
\section{Conclusions \label{sec:5}}

In this work, we considered the soft drop groomed jet radius $\theta_g=R_g/R$ at next-to-leading logarithmic accuracy. The radius of a soft drop groomed jet is one of the key observables characterizing the impact of grooming on a jet and is calculable in perturbative QCD. It is defined as the opening angle of the splitting that satisfies the soft drop grooming condition and is related to the active area of the groomed jet. The extension of the calculation beyond leading-logarithmic accuracy required us to study the nontrivial all order structure of non-global logarithms which are affected by clustering constraints due to the use of the C/A algorithm. In addition, Abelian clustering logarithms need to be taken into account. An important ingredient to understand the factorization structure is the equivalence between the soft drop groomed radius measurement and a jet veto between the boundaries of the groomed and ungroomed jet. Within SCET we established a factorization theorem which allows for the resummation of logarithms of $\theta_g$, the jet radius $R$ and the soft drop parameter $z_{\rm cut}$ 
at NLL. We performed an explicit calculation of the non-global and Abelian clustering logarithms at fixed order. The all order resummation at leading logarithmic accuracy within the large-$N_c$ approximation was achieved by making use of a suitably designed Monte Carlo algorithm. We performed numerical calculations and compared our results to Pythia 8 simulations for LHC kinematics and found very good agreement. From these findings, we concluded that the soft drop groomed radius is under good perturbative control as the overall impact of nonperturbative effects, mainly due to hadronization, appears to be relatively small for LHC kinematics. Numerical predictions for the STAR experiments at RHIC are also provided.

Our results allow for precision comparisons to data from the LHC and RHIC which will further improve our understanding of soft drop groomed jet substructure observables. The formalism developed in this work can be systematically extended beyond NLL accuracy and matched to fixed order calculations. In addition, it will be interesting to systematically investigate nonperturbative effects. Applications in heavy-ion collisions will further extend the use of the calculations performed in this work.

\section*{Acknowledgments}

We thank T.~Becher, R.~Elayavalli, P.~Jacobs, A.~Larkoski, M.~LeBlanc, Y.-J. Lee, S.~Marzani, J.~Mulligan, B.~Nachman, M.~Ploskon, J.~Roloff, D.-Y.~Shao, G.~Soyez and F.~Yuan for helpful discussions. Z.K. is supported by the National Science Foundation under Grant No.~PHY-1720486. K.L. is supported by the National Science Foundation under Grants No.~PHY-1316617 and No.~PHY-1620628. X.L. is supported by the National Natural Science Foundation of China under Grant No.~11775023 and the Fundamental Research Funds for the Central Universities. D.N is supported from Department of Energy contract DE-AC52-06NA25396 at LANL and through the LANL/LDRD Program via a Feynman Distinguished Fellowship. F.R. is supported by the Department of Energy under Contract No.~DE-AC0205CH11231, the LDRD Program of LBNL and by the National Science Foundation under Grant No. ACI-1550228 within the JETSCAPE Collaboration.

\appendix

\section{Collinear-soft emissions at NNLO~\label{sec:NNLO}}

Here we consider the phase space for collinear-soft emissions at NNLO as a non-trivial example to check and illustrate the equivalence between the soft drop groomed $R_g$ measurement and the jet veto constraint. We first introduce the notation for two collinear-soft partons $i$ and $j$ with momenta $k_{i}$ and $k_j$ as
\bea
& \theta_{i\!j} \equiv \Delta R_{ij} \,, \quad \quad
k_{ij} \equiv k_i + k_j\,,  \nn \\
& \Theta_{i,p} \equiv \Theta\left(  \frac{p_{Ti}}{p_{TJ}} > z_{\rm cut} \left(  \frac{ \Delta R_{i,J} }{R} \right)^\beta   \right) \,,  \\
& \Theta_{i,f} \equiv \Theta\left(  \frac{p_{Ti}}{p_{TJ}} <  z_{\rm cut} \left(  \frac{ \Delta R_{i,J} }{R} \right)^\beta   \right) = 1 - \Theta_{i,p} \,, \nn  
\eea
where $J$ denotes the eikonal direction set by the collinear mode. We further write the soft drop groomed radius $R_g$ phase space for $1$ collinear-soft emission as 
\bea\label{eq:M1def}
{\cal M}_1(k_i) \equiv \Theta(R_g - \theta_{iJ})\Theta_{i,p} + \Theta_{i,f} \,, 
\eea
which is equivalent to  
\bea
{\cal M}_1(k_i) \equiv \Theta(R_g - \theta_{iJ})  + \Theta(\theta_{iJ} - R_g) \Theta_{i,f} \,.
\eea
The first term indicates that an emission $i$ is kept as long as it is within the cone set by $R_g$, whereas it is vetoed if it is outside. This is the usual veto operation and manifests the equivalence at NLO. For future use, we also note that 
\bea\label{eq:M1bardef}
1 - {\cal M}_1(k_i) = \Theta(\theta_{iJ} - R_g  )  \Theta_{i,p} \,. 
\eea
\begin{figure}[t]\centering
\includegraphics[width=3.3in]{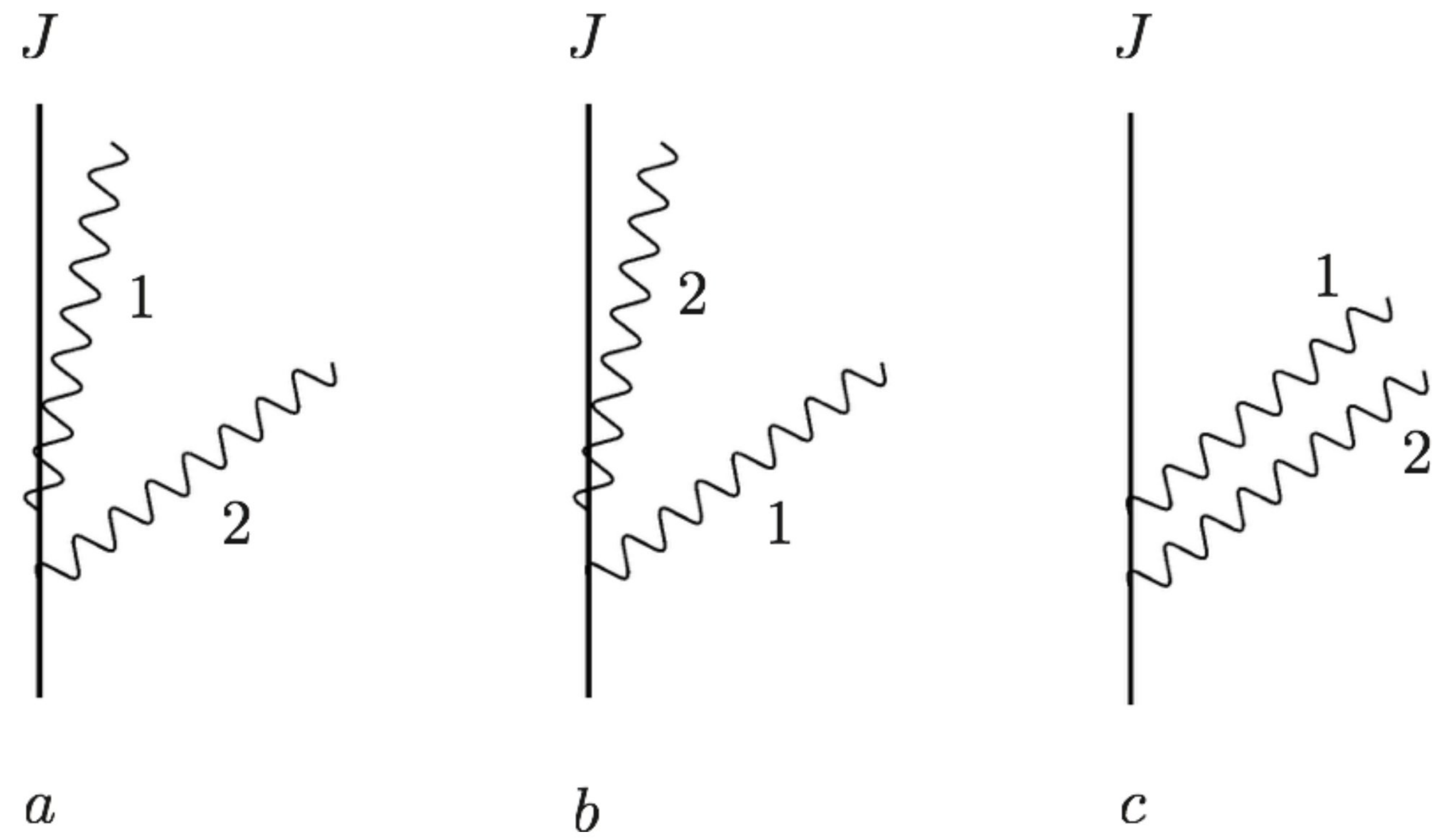} 
\caption{Configurations of collinear-soft emissions at NNLO. Either emissions 1 or 2 are closest to the eikonal direction $J$ ($a$, $b$) or 1 \& 2 get clustered into one branch first and are then combined with $J$ ($c$).~\label{fig:three}}
\end{figure}
At NNLO, we need to consider the three configurations shown in Fig.~\ref{fig:three}. We start with configuration $a$. In this case we can write the phase space measure as
\bea
{\cal M}_{2}^a   \,= &\,\Theta(\theta_{2J} - \theta_{1J} )\, \Theta(\theta_{12} - \theta_{1J})\,\left[\,\Theta_{2,p}
\Theta({R_g} - \theta_{2J} ) 
+ \Theta_{2,f} \Big(
\Theta(R_g - \theta_{1J} ) \Theta_{1,p} + \Theta_{1,f}
\Big)\,
\right] \nn \\
  = &\,\Theta(\theta_{2J} - \theta_{1J} ) \,\Theta(\theta_{12} - \theta_{1J})\, \left[\,\Theta_{2,p}
\Theta({R_g} - \theta_{2J} ) 
+ \Theta_{2,f} {\cal M}_1(k_1)\,
\right] \,,
\eea
where the overall factor indicates that $\theta_{1J}$ is the smallest distance. When declustering the jet, we thus first check emission 2 against the soft drop criterion. If $2$ passes, it has to be within $R_g$ by construction and the algorithm terminates (first term in square brackets). If emission 2 fails, we continue to check whether emission 1 passes the soft drop criterion (second term in square brackets). To proceed, we write ${\cal M}_2^a$ as follows
\bea
{\cal M}_2^a = &\, \Theta(\theta_{2J} - \theta_{1J} ) \Theta(\theta_{12} - \theta_{1J}) \bigg[\Theta_{2,p}
\Theta({R_g} - \theta_{2J} ) +   {\cal M}_{1}(k_2)  {\cal M}_1(k_1) 
\nn \\
&-  \Theta(R_g - \theta_{2J})\Theta_{2,p} \,  {\cal M}_1(k_1)  \bigg]
\eea
where we used the definition of ${\cal M}_1(k_2)$ in eq.~(\ref{eq:M1def}). Now we combine the first and the third term and use eq.~(\ref{eq:M1bardef}) to obtain
\bea
{\cal M}_2^a \,= \, \Theta(\theta_{2J} - \theta_{1J} )\,\Theta(\theta_{12} - \theta_{1J})  
 \,\left[\,
\Theta({R_g} - \theta_{2J} ) \, \Theta(\theta_{1J} - R_g  ) \, \Theta_{1,p} \, \Theta_{2,p}
+  {\cal M}_1(k_1)  \,{\cal M}_{1}(k_2)\, \right] \,. 
\eea
The first term vanishes since there is a contradiction between the different conditions requiring $\theta_{1J}$ to be the smallest distance as well as $\theta_{1J} > R_g$ and $R_g > \theta_{2J}$. Therefore, we find for the configuration $a$ the result
\bea
{\cal M}_2^a \,=\, \Theta(\theta_{2J} - \theta_{1J} ) \,\Theta(\theta_{12} - \theta_{1J})\,
   {\cal M}_1(k_1)\,{\cal M}_{1}(k_2)
 \,.
\eea
For configuration $b$ the same arguments apply. We can thus write the measurement functions for $a+b$ as
\bea
{\cal M}_2^a + {\cal M}_2^b \, = \,\left[\,
1 -    \Theta(\theta_{1J} - \theta_{12} ) \, \Theta(\theta_{2J} - \theta_{12})\, \right]  \,{\cal M}_1(k_1)\,{\cal M}_{1}(k_2) \,.
\eea
This states that we separately veto emissions 1 and 2 as long as $\theta_{12}$ is not the smallest overall distance in which case they will be combined into a single branch first. Lastly, we consider the configuration $c$ in Fig.~\ref{fig:three} which can be written as
\bea
{\cal M}_2^c \,=\,  \Theta(\theta_{1J} - \theta_{12} ) \,\Theta(\theta_{2J} - \theta_{12})
\,{\cal M}_1(k_{12}) \,.
\eea
If the distance $\theta_{12}$ is the smallest distance, the emissions 1 and 2 will be combined first. In the declustering procedure, the branch containing both emissions will be checked againnst the soft drop criterion. Adding up all three configurations, we find 
\bea
{\cal M}_2^a + {\cal M}_2^b + {\cal M}_2^c \, =&\, \left[\,
1 -    \Theta(\theta_{1J} - \theta_{12} )\, \Theta(\theta_{2J} - \theta_{12})  
\,\right] \,{\cal M}_1(k_1) \,  {\cal M}_{1}(k_2)  \nn \\
& +  \,\Theta(\theta_{1J} - \theta_{12} )\, \Theta(\theta_{2J} - \theta_{12}) \,{\cal M}_1(k_{12}) 
\,,
\eea
which shows the equivalence discussed in the main text at NNLO. The extension beyond NNLO can be achieved in a similar way. For instance, in the case of 3 emissions, the only extra configuration one needs to consider is when neither $2$ nor $3$ emissions out of the 3 are clustered first. Otherwise, up to clustering, it is equivalent to the $1$- or $2$-emission case discussed here in detail. Following similar steps and arguments as we showed for configuration $a$ of the $2$-emission case, one reaches again the equivalence.   

At NNLO, we can also have $2$ collinear emissions $J_1$ and $J_2$ along with $1$ collinear-soft parton $k_1$. The measurement function is readily found to be given by
\bea
{\cal M}_{J_1J_2} \,= \,& \,
{\cal M}_1(k_1) 
+ \left[\,
\Theta(\theta_{1 J_2}
-\theta_{1 J_1})\,
\Theta(\theta_{J_1J_2} 
- \theta_{1J_1}) 
+  J_1\leftrightarrow J_2
\,\right]
\,\left[\,
1 - {\cal M}_1(k_1) 
\,\right] \,,
\eea
which again is equivalent to the jet veto operation using the C/A algorithm. That is to say, when the soft parton is combined with the collinear radiation $J_1$ or $J_2$, it will be kept. Otherwise, it will be vetoed when it is outside of the region set by $R_g$.

\medskip

\bibliographystyle{JHEP}
\bibliography{bibliography}

\end{document}